\documentclass[aps,twocolumn,showkeys,superscriptaddress,10pt]{revtex4-2}
\usepackage{amsmath,amssymb,graphicx}
\usepackage[percent]{overpic}
\usepackage{bm,subfigure,braket,float,pstricks}
\usepackage[colorlinks={true}]{hyperref}
\hypersetup{colorlinks=true,linkcolor=red,citecolor=magenta,urlcolor=blue}
\usepackage{xcolor}

\usepackage[left=0.3in, right=0.3in, top=0.7in, bottom=0.7in]{geometry}
\usepackage{lipsum}
\usepackage{orcidlink}

\begin{document}
\title{Gravitational cat states as a resource for quantum information processing}
\author{Atta ur Rahman \orcidlink{0000-0001-7058-5671}}
\affiliation{School of Physical Sciences, University of Chinese Academy of Sciences, Yuquan Road 19A, Beijing 100049, China}
\author{Ao-Xiang Liu}
\affiliation{School of Physical Sciences, University of Chinese Academy of Sciences, Yuquan Road 19A, Beijing 100049, China}
\author{Saeed Haddadi \orcidlink{0000-0002-1596-0763}}
\email{haddadi@ipm.ir}
\affiliation{School of Particles and Accelerators, Institute for Research in Fundamental Sciences (IPM), P.O. Box 19395-5531, Tehran, Iran}
\author{Cong-Feng Qiao \orcidlink{0000-0002-9174-7307}}
\email{qiaocf@ucas.ac.cn}
\affiliation{School of Physical Sciences, University of Chinese Academy of Sciences, Yuquan Road 19A, Beijing 100049, China}

\begin{center}
\begin{abstract}
\vspace{1ex}
We investigate how resourceful gravitational cat states are to preserve quantum correlations. In this regard, we explore the dynamics of gravitational cat states under different situations such as thermal, classical stochastic, general decaying, and power-law noisy fields. In particular, the one-way steerability, Bell non-locality, entanglement, and purity in two qubits are our main focus.  We also address the weak measurement protocol on the dynamics of quantum correlations and purity of the state.  Our results show that the gravitational cat states have a reliable and better capacity to preserve quantum correlations and remain one of the good resources for the deployment of quantum information processing protocols. Additionally, two independent channels are also employed and it is observed that only the weaker coupling regimes are effective in preserving quantum correlations.  Notably, in terms of non-Markovian dynamics implication, quantum correlations are found to be longer preserved because of the information feedback phenomenon between the system and environment. Finally, we present a brief analysis to extend our gravitational model to include the electrostatic notion, providing insight into the key differences between the considered configurations.
\end{abstract}
\end{center}

\maketitle

\section{Introduction}
In order to analyze, preserve, and process information in ways that are not possible with conventional systems, quantum information and computation use the superposition of quantum subsystems. Such states have the potential to solve complex issues faster and with ease, such as the computation of problems involving quantum many-body systems \cite{1}. Furthermore, quantum systems are characterized by certain correlations such as quantum entanglement that cannot be defined using any classical means \cite{2}, Bell non-locality \cite{3}, quantum discord \cite{4}, etc. A key idea in quantum physics as well as in existing and upcoming quantum technologies is entanglement, which appears at incredibly small and subatomic levels, comparable to other non-classical characteristics. When two or more particles get entangled, they continue to influence the measurement of each other even if isolated by a vast distance. Quantum entanglement may be used for communication by making use of the unique correlations that entangle qubits, as given in Ref. \cite{5}. Interestingly, the instantaneous information transmission over very long ranges is made possible by entangled qubits \cite{6, 7}. Entangled states, for instance, can be a resource for building efficient quantum computers, which may lead to the development of modern computing protocols and associated technologies such as machine learning \cite{8} and drug discovery \cite{9}. Quantum communication can be made extremely secure by utilizing entanglement \cite{10} as well as quantum networks with extremely long lengths \cite{11}.
Perhaps soon, thanks to entanglement, a fully functional quantum internet that allows everyone to communicate as a service will be available to everyone \cite{12}.
\par
Einstein-Podolsky-Rosen (EPR) steering is a quantum occurrence that Schr\"{o}dinger attempted to formalize in 1935 \cite{13}. By selecting particular local measurements, one witness may be able to remotely manipulate or change the quantum state of some other participant \cite{14}. The presence of entanglement, EPR steering, and Bell non-locality was demonstrated in the ordering of the quantum non-separable states \cite{13, 14}. As a result, the Bell non-locality exists in steered states, which are a subclass of entangled states  \cite{13}. A single-photon Bell state has been found ideal for quantum steering because it has greater loss endurance compared to other quantum correlation criteria \cite{15}. The inequality of the continuous variable EPR steering has vanished with the non-degenerate optically controlled oscillator \cite{16}. Experimental research has been done to determine whether it is possible to observe EPR steering in two entangled photon optical states \cite{17}. Additionally, quantum steering and associated steerability for a range of non-local systems were developed by addressing steering inequalities, for instance, in the cases of bipartite two-qubit X-state \cite{18}, accelerated hybrid (qubit-qutrit) systems \cite{19}, multi-qubit states when influenced by Unruh effect \cite{20}, two qubits under classical dephasing channels \cite{rahmanadp}, and two-qubit spin chains \cite{21}. In the above papers, it has been demonstrated that the degree of steerability between the sub-components of a quantum state changes with varying dimensions, coupled fields, and associated phenomena. Hence, this motivates the prospects for exploring EPR steering for different quantum systems assumed in different situations and protocols.
\par
Bell non-locality is a fundamental quantum feature that is unsurpassed by any classical model and functions as a kind of stronger quantum correlation. Moreover, Bell non-locality is characterized by Bell inequality and occurs when a bipartite quantum state deviates from a specific Bell inequality and local measurement results cannot be explained by classical random distributions over probability domains \cite{23}. Bell non-locality is strongly indicated by the violation of the Bell inequality, particularly the Bell-CHSH inequality, both theoretically and empirically \cite{24}.
It is now recognized as a valuable resource required for both quantum computation and communication. Bell inequality violations could be used to determine entanglement, shedding new light on how Bell inequalities are used in practice \cite{26}.
Hence, the relationship between entanglement and non-locality has received a great deal of attention \cite{26-2}.   Accepting local hidden variable models is crucial in a quantum state devoid of entanglement. Therefore, it is important to realize the relationship between entanglement and Bell non-locality, as well as their dynamics and preservation in different open quantum states, which may lead us to enhance several quantum technological protocols.
\par
In a quantum interpretation of the matter, a stationary heavy particle may be assumed to be in a Schr\"{o}dinger cat state or in a superposition of two physically distinct positions \cite{27}. These states are referred to as gravitating entities and are known as gravitational cats (gravcats) \cite{28-1,28,Rojas2023,HaddadiCTP2024,HaddadiEPJC2024}. It is fundamentally important to comprehend how these states behave, particularly, in the disciplines of gravitational non-local physics and macroscopic non-local events. Therefore, gravcats can be used to assess the resiliency of quantum systems for practical quantum information deployment. In comparison to the plethora of quantum gravity theories that claim to explain current physics' minimal energy scoring system \cite{29, 30, 31}, even in the available weak-gravitational and non-relativistic regimes \cite{28, 32}, experiments and recommendations about the interaction of gravitation and quantum matter are generally insufficient. In particular, the investigation of gravcats, as a resource, in terms of open quantum systems dynamics and quantum information science needs to be assessed in detail.
\par
In this article, we treat a general gravcat as an open quantum system when coupled with various types of external fields, for example, thermal, classical (with and without decay) and a specific noise-assisted dephasing field. As various parameters influence the dynamics of realistic open quantum systems, we take into account an external thermal field governed by Gibbs density operator. Besides, we assume that a classical field will be employed in collaboration with the thermal field in three different schemes. In the first case, we combine the thermal field with the classical fluctuating field without having any decaying effects. The thermal field is then combined with a local field characterized by a general decaying function in the second place, while finally, a classical field with dephasing caused by a specific noise is imposed on the thermal field. The motive behind using such a joint scheme is to investigate the dynamics of quantum resources in gravcats under the influence of various external situations. For example, Gibbs density operator can let us know about the impact of temperature, which is important. However, this gravcat-thermal field coupling cannot provide any details about the time evolution and external dephasing effects, which are part of realistic quantum computing and will be an ideal situation to discuss. These factors, such as the time evolution of open quantum systems and external dephasing effects can be investigated by including external classical channels. It is expected to provide an optimal channel for the higher possible preservation of quantum correlations in the considered gravcat system under the influence of various external impacts. Furthermore, the dynamics of the steerability, non-locality, entanglement, and mixedness are done by utilizing the one-way steering, Bell non-locality, concurrence, and purity measures.
\par
This paper is organized as follows. In Sec. \ref{model}, we provide the details of the considered physical model, quantum correlation measures, and main results. In Sec. \ref{Significance}, we express the significance and experimental prospects of our study. Finally, in Sec. \ref{conclusion}, we conclude our work.

\begin{figure}[t]
	\begin{center}		\includegraphics[width=0.48\textwidth]{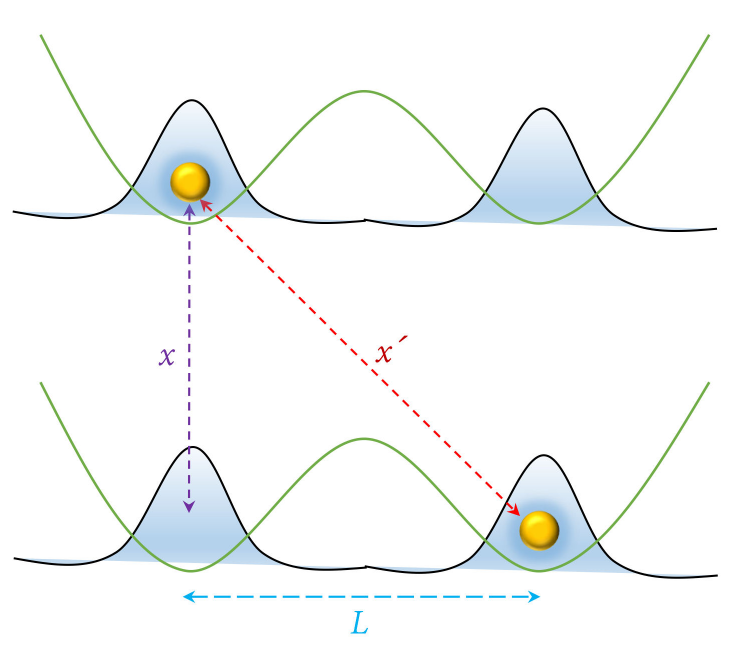}
		\end{center}
\caption{Schematic representation of the gravcats model. Each symmetric double-well potential is located along a distinct axis, while two axes are parallel at a distance $x =\sqrt{{x^{\prime}}^2-L^2}$.} \label{fig0}
\end{figure}

\section{Dynamics of gravcat state}\label{model}

\subsection{Hamiltonian and thermal interaction}
We assume a particle confined in one dimension of mass $m$ having the Hamiltonian $\hat{H}_P=\frac{\hat{p}^2}{2m}+V(\hat{s})$ where $V(s)$ is the symmetric double-well potential with local minima at $s=\pm L/2$ \cite{28}.  Let us assume that $V(s)$ is even, therefore, all eigenstates of Hamiltonian are parity definite. Note that the least energy eigenstate $\vert g \rangle$ is parity symmetric and the first excited state $\vert e \rangle$ is parity-antisymmetric. We suppose the energy difference between the ground and excited states $\vert g \rangle$ and $\vert e \rangle$ is denoted by $\omega$, then we define the vector superposition state as
$\vert \pm \rangle=(\vert g \rangle+\vert e \rangle)/\sqrt{2}$.
\par
Now, let us consider the Hamiltonian of the physical model describing the gravitational interaction in the non-relativistic limit between the two-qubit gravcats state, given by \cite{28}
\begin{align}
\mathcal{H} = \frac{\omega}{2}\left(\sigma_{z}\otimes \mathbb{I}+\mathbb{I}\otimes \sigma_{z}\right)-\gamma (\sigma_{x}\otimes \sigma_{x}), \label{hamiltonian}
\end{align}
where $\sigma_{x}$ and $\sigma_{z}$ denote the Pauli spin-$1/2$ operators, $\mathbb{I}$ is the $2 \times 2$ identity matrix, and $\gamma:=G\frac{m^{2}}{2}\left(\frac{1}{x}- \frac{1}{x'}\right)$ regulates the gravitational interaction coupling strength between the two massive gravcat states. Here, $G$ is the universal gravitational constant, and $x$ and $x'$ are the relative distances between the two particles \cite{Rojas2023}, as depicted in Fig. \ref{fig0}.

Next, we consider that the gravcats are exposed to a thermal reservoir at an equilibrium temperature $T$, and the related thermal density operator can be written as
\begin{equation}
\rho(T)=\frac{1}{Z}\exp\left(  -\mathcal{H}/k_{B} T\right),\label{thermal density matrix1}
\end{equation}
where $Z=\text{Tr}\left[\exp\left(-\mathcal{H}/k_{B} T\right)\right]$ is the partition function, $T$ is the absolute temperature, and $k_{B}$ is the Boltzmann constant (considered $k_{B}=1$ for simplicity).

Finally, the explicit form of the thermal density matrix \eqref{thermal density matrix1} based on Eq. \eqref{hamiltonian} takes the form
\begin{equation}
\rho_T=\left[\begin{array}{cccc}
\rho_{11} & 0 & 0 & \rho_{14}\\
0 & \rho_{22} & \rho_{23} & 0\\
0 & \rho_{32} & \rho_{33} & 0\\
\rho_{41} & 0 & 0 & \rho_{44}
\end{array}\right],\label{eq:rho-mat}
\end{equation}
where the entries are
\begin{align}
\rho_{11}= & \frac{1}{Z}\left[\cosh \left(\mathcal{K}/T\right)-\frac{\omega  \sinh \left(\mathcal{K}/T\right)}{\mathcal{K}}\right],\nonumber \\
\rho_{14}=&\rho_{41}=  \frac{\gamma  \sinh \left(\mathcal{K}/T\right)}{Z \mathcal{K}},\nonumber \\
\rho_{22}=&\rho_{33}=  \frac{\cosh \left(\gamma /T\right)}{Z},\nonumber \\
\rho_{23}=&\rho_{32}=  \frac{\sinh \left(\gamma/ T\right)}{Z}, \nonumber \\ \rho_{44}= & \frac{1}{Z}\left[\cosh \left(\mathcal{K}/T\right)+\frac{\omega  \sinh \left(\mathcal{K}/T\right)}{\mathcal{K}}\right],
\label{thermal density matrix}
\end{align}
with $Z=2 \left[\cosh \left(\mathcal{K}/T\right)+\cosh \left(\gamma /T\right)\right]$ and $\mathcal{K}=\sqrt{\gamma ^2+\omega ^2}$.

\begin{figure*}[!t]
	\begin{center}
		\includegraphics[width=0.45\textwidth, height=145px]{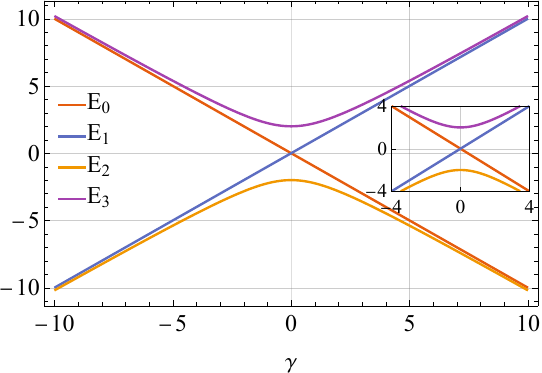}
		\put(-200,148){($ a $)}\
		\includegraphics[width=0.45\textwidth, height=145px]{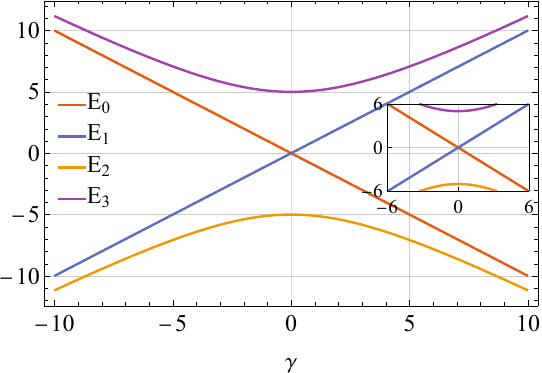}
		\put(-200,148){($ b $)}\\
		\includegraphics[width=0.45\textwidth, height=145px]{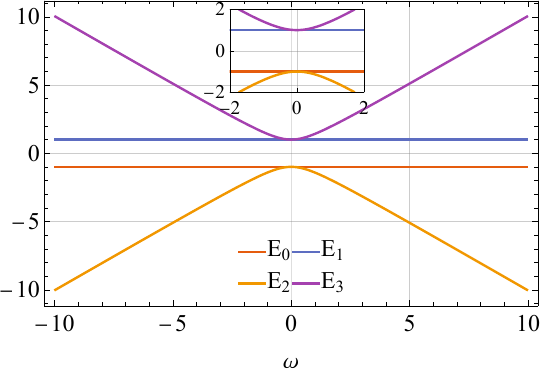}
		\put(-200,148){($ c $)}\
		\includegraphics[width=0.45\textwidth, height=145px]{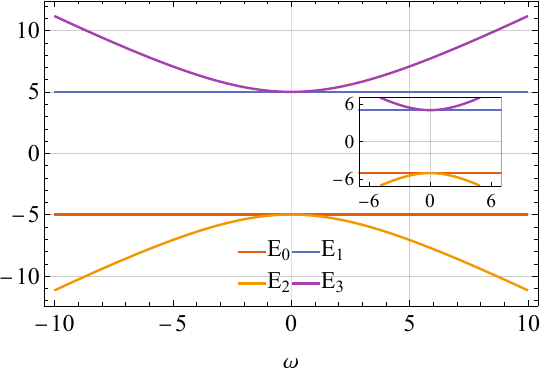}
		\put(-200,148){($ d $)}
		\end{center}
\caption{ Four eigenvalues of the Hamiltonian versus $\gamma$ and $\omega$ when (a) $\omega=1$, (b) $\omega=5$,  (c) $\gamma =1$, and (d) $\gamma =5$.} \label{QPT}
\end{figure*}

In Fig. \ref{QPT}, we plot the eigenvalues of the Hamiltonian \eqref{hamiltonian}  for different $\omega$ values against $\gamma$ (top row) and so the opposite (bottom row). Explicitly, the eigenvalues of the system are as $E_0=-\gamma,$ $E_1=\gamma,$ $E_2=-\mathcal{K}$, and $E_3=\mathcal{K}$.  Besides,  the corresponding eigenvectors read $|\psi_{0}\rangle= \frac{1}{\sqrt{2}}\left\{|01\rangle+|10\rangle\right\}$, $|\psi_{1}\rangle= \frac{1}{\sqrt{2}}\left\{-|01\rangle+|10\rangle\right\}$,
$|\psi_{2}\rangle = k_- \left\{-(\omega -\mathcal{K})|00\rangle+\gamma |11\rangle \right\}$, and $|\psi_{3}\rangle=k_+ \left\{-(\omega +\mathcal{K})|00\rangle+\gamma |11\rangle \right\}$,  where  $k_\pm=1/\sqrt{|\omega\pm \mathcal{K}|^2+\gamma^2}$.
From Fig. \ref{QPT}, one can notice a crossover between $E_0$ and $E_1$ at critical point $\gamma=0$.
At this point, $E_2$ and $E_3$ become diverge from  $E_0$ and $E_1$. On the contrary, for the increasing values of $|\gamma|$, both the mentioned pairs seem to converge, showing a sign of quantum phase transitions. In Fig. \ref{QPT}(b), we only see a divergence of the pair $E_2$ and $E_3$ from $E_0$ and $E_1$ at the critical point $\gamma=0$. Besides, the slopes of the mentioned pairs also seem to diverge even at higher $|\gamma|$ values. In Fig. \ref{QPT}(c), we set $\gamma=1$. It is clear that $E_0$ and $E_1$ are independent of parameter $\omega$, however, $E_2$ and $E_3$ alter as $\omega$ changes.
Note that when $\omega \rightarrow 0$,  we observe the crossovers between $E_3$ and $E_1$ as well as $E_2$ and $E_0$.
Similar results are explored in Fig. \ref{QPT}(d), however, the space between $E_0$ and $E_1$ becomes larger and the crossovers of $E_3$ with $E_1$ and $E_2$ with $E_0$ occur quickly.

\subsection{Classical stochastic field}\label{classical}
Afterwards, we are interested in including what happens to the initially encoded quantum correlations in the gravcat state with time. The motive behind using such a concept is that the previous mathematical machinery given up to Eq. \eqref{eq:rho-mat} lacks the time parameters, therefore, we will be unable to judge the qualitative and quantitative analysis of the dynamics, preservation, and dephasing effects of quantum correlations in the gravcats with time. For this reason, we propose doing this by exposing the current physical model to classical fluctuating fields. {  Note that the ideas of classical and stochastic fields have been discussed in experimental aspects in Refs. \cite{d1, d2}.} Therefore, let us consider a two-qubit Hamiltonian, given as \cite{36}
\begin{equation}
H(t)={H}_i (t)\otimes \mathbb{I}_j \otimes+\mathbb{I}_i \otimes{H}_j(t),\label{hmm}
\end{equation}
where
\begin{equation}
H_k (t)=E \mathbb{I}+\lambda \delta_k (t) \sigma_z=\left[
\begin{array}{cc}
 E+\delta_k  (t)\lambda  & 0  \\[0.2cm]
 0 & E-\delta_k  (t)\lambda   \\
\end{array}
\right],\label{inde ham}
\end{equation}
is the individual Hamiltonian of the sub-quantum system $k$ (with $k \in\{i, j\}$), where $E$ is the energy of the single qubit,  $\lambda$ is the intensity of the system–environment coupling, the stochastic parameter $\delta_k  (t)$ controls the fluctuation of the classical field and flips between $+1$ and $-1$, and $\sigma_z$ is the spin-$1/2$ Pauli operator.

We use the time unitary operator for the time-evolution of the two-qubit system \eqref{hmm} as (setting $\hbar=1$)
\begin{equation}
U(t)=\exp \left\{-i \int^{t}_{0} H (t^{\prime})dt^{\prime} \right\}.\label{ut}
\end{equation}

Then, we assume a common classical channel, i.e., exposing the two-qubit system to a single classical channel which is done by setting $\delta_i(t)=\delta_j(t)=\delta$. Next, for obtaining the time-evolved state of the initial thermal density matrix  $\rho_T\equiv\rho(0, T)$  \eqref{eq:rho-mat}, we employ the expression
\begin{equation}
\rho(t, T)=U(t)\rho(0, T) U^{\dagger}(t).\label{time evolved density matrix}
\end{equation}

Finally, the explicit form of the thermal-time-evolved density matrix can be written as
\begin{align}
\rho(t, T)=\left[
\begin{array}{cccc}
 \rho_{11} & 0 & 0 &  e^{-4 i \delta \lambda  t} \rho_{14}\\[0.2cm]
 0 & \rho_{22} & \rho_{23} & 0 \\[0.2cm]
 0 & \rho_{23} & \rho_{22} & 0 \\[0.2cm]
 \rho_{14} e^{4 i \delta \lambda  t} & 0 & 0 & \rho_{44} \\
\end{array}
\right].\label{rho-t}
\end{align}

Let us examine the impact of the joint classical and thermal field on the dynamics of the entangled gravcat system prepared in the state $\rho(t, T)$. Here, we consider an entanglement witness (EW) operation to detect entanglement in the state and the associated dynamical map. The EW operator mathematically can be written as

\begin{equation}
\textmd{EW}:=-{\rm Tr}[\mathcal{W}_{op}\rho(t, T)]\label{ewww}
\end{equation} 
where
$\mathcal{W}_{op}:=\frac{1}{2}\mathbb{I}_{4 \times 4}-\rho(0, T)$ is our considered EW operator. Notice, $\textmd{Tr}[\mathcal{W}_{op} \rho(t, T)] <0$ validates the presence of entanglement, however, $\textmd{Tr}[\mathcal{W}_{op} \rho(t, T)]>0$ does not restrict us from the state $\rho(t, T)$ to be entangled \cite{EW}.
Therefore, we obtain an analytic expression of EW for our gravcat state as follows

\begin{equation}
\textmd{EW}=\frac{\exp\left[2 (\mathcal{K}+\gamma )/T\right] (\Upsilon_1-4\Upsilon_2+\Upsilon_3+2 \Upsilon_4)}{\mathcal{K}^2 \left(\exp[\mathcal{K}/T]+\exp[\gamma /T]\right)^2 \left(\exp[(\mathcal{K}+\gamma)/T]+1\right)^2}, \label{EW-reults}
\end{equation}
where
\begin{align*}
\Upsilon_1=&-\gamma ^2+\mathcal{K}^2 \cosh \left(2 \gamma /T\right)-2 \omega ^2, \\
\Upsilon_2=&\mathcal{K}^2 \cosh \left(\gamma /T\right) \cosh \left(\mathcal{K}/T\right),\\
\Upsilon_3=&\omega ^2 \cosh \left(2 \mathcal{K}/T\right),\\
\Upsilon_4=& \gamma ^2 \cos (4 \delta \lambda  t) \sinh ^2\left(\mathcal{K}/T\right).
\end{align*}

\begin{figure*}[!t]
	\begin{center}
		\includegraphics[width=0.32\textwidth, height=145px]{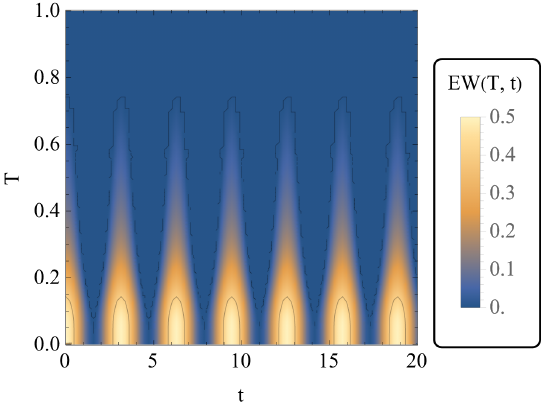}
		\put(-145,148){($ a $)}\quad
		\includegraphics[width=0.32\textwidth, height=145px]{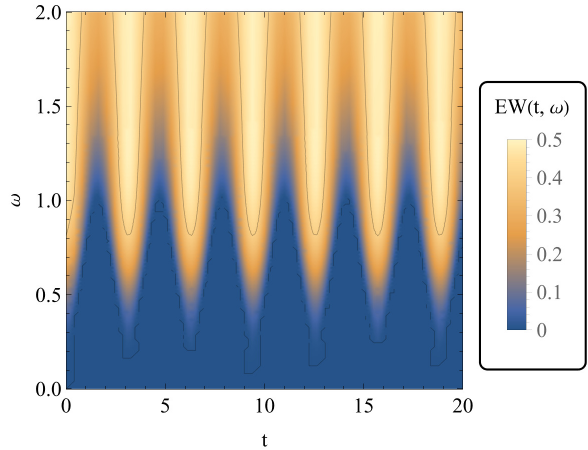}
		\put(-145,148){($ b $)}\quad
		\includegraphics[width=0.32\textwidth, height=145px]{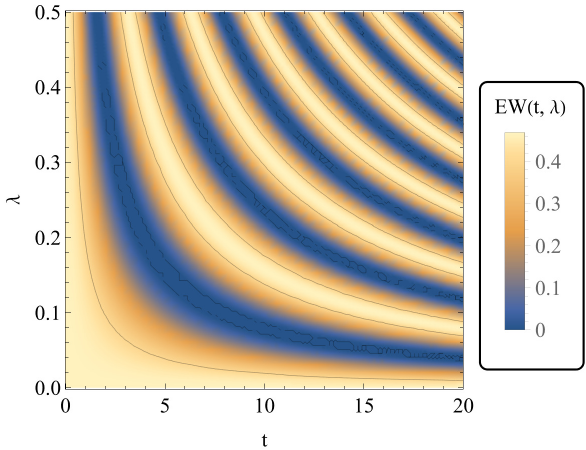}
		\put(-145,148){($ c $)}\\
		\includegraphics[width=0.32\textwidth, height=145px]{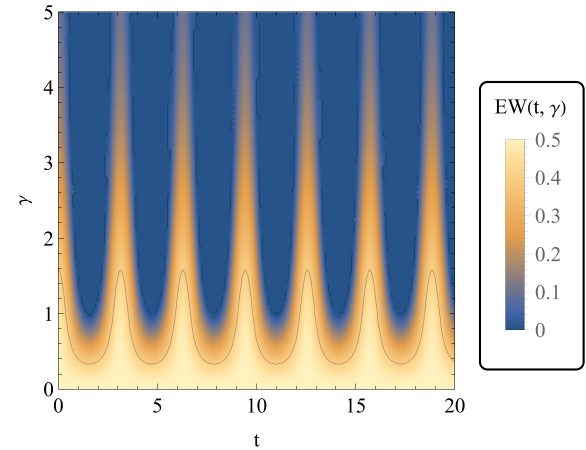}
		\put(-145,148){($ d $)}\quad
		\includegraphics[width=0.32\textwidth, height=145px]{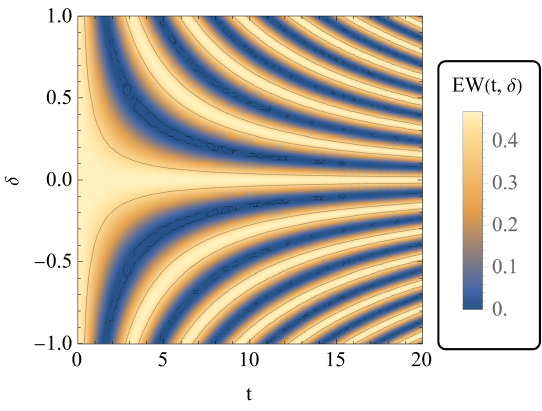}
		\put(-140,148){($ e $)}
		\end{center}
\caption{Dynamics of EW for the two-qubit gravcat state prepared in the state $\rho(t, T)$ given in Eq. \eqref{rho-t} influenced by a classical stochastic field when  (a) $\delta=\gamma=\omega=1,~\lambda=0.5$, (b) $\delta=\gamma=1,~\lambda=0.5,~T =0.1$, (c) $\delta=\gamma=\omega=1,~T =0.1$, (d) $\delta=\omega=1,~\lambda=0.5,~T =0.1$, and (e) $\gamma=\omega=1,~ \lambda=0.5,~T =0.1$.} \label{EW}
\end{figure*}

Fig. \ref{EW} deals with the detection of the entanglement in the two-qubit gravcat state and associated dynamical map under different circumstances. In the first case, Fig. \ref{EW}(a), the impact of temperature with time on the entanglement of the gravcat state is shown and it is found that for the lower temperature values, the system is detected in the maximally entangled range. For the increasing temperature, entanglement in the system seems lost, specifically, for $T>0.7$ the state becomes disentangled.  In the second plot [Fig. \ref{EW}(b)], we presented the impact of $\omega$ which has been found a crucial parameter for the robustness of the entanglement in the gravcat state. As seen for lower $\omega$ values, the state remains disentangled and the opposite occurs for the higher $\omega$ values. The effect of the system and classical field coupling constant $\lambda$ has been observed in Fig. \ref{EW}(c) to drive the conversion of the entangled state into the free state and vice versa. As can be seen that for increasing $\lambda$ values, the oscillations in the dynamical map of the EW measure increase. Note that for the higher $\lambda$ values, the time of entanglement becomes narrower and for lower $\lambda$ values, the system remains entangled for a longer time as seen per oscillation. In Fig. \ref{EW}(d), the influence of $\gamma$ has been witnessed opposite to that observed for $\omega$ in Fig. \ref{EW}(b).  Finally, in Fig. \ref{EW}(e), the impact of stochastic parameter $\delta$ has been observed which has the flipping range $-1 \leq \delta \leq 1$. We find that for the increasing $|\delta|$, the oscillations in the dynamical maps of the system increase and that both the positive and negative values of $\delta$ similarly affect the dynamical map of the EW. Notice, the EW measure faces sudden deaths and births over time. This suggests the inter-conversion of the entangled state into a free state and vice versa. Besides, it also shows that the system and fields strongly exchange information between themselves.  It is worth noting that there is no imaginary part in the eigenvalues obtained from the thermal Hamiltonian given in Eq. \eqref{hamiltonian}. Hence, the oscillating behavior is due to the Hamiltonian obtained in Eq. \eqref{hmm} for the classical field. 

{  In terms of the physical interpretation, we find the temperature of the system plays a key role; at low temperatures, the system remains highly entangled. The $\omega$ parameter acts as a stabilizing force for entanglement where higher $\omega$ strengthens it, making the system more robust against disturbances. The coupling constant $\lambda$ directly affects entanglement fluctuation and randomness over time. As $\lambda$ increases, the system experiences more rapid oscillations between entangled and disentangled states, though these periods of entanglement become shorter. In contrast, the parameter $\gamma$ influences the system in the opposite way to $\omega$, further emphasizing how sensitive entanglement is to the interplay of different factors. The stochastic parameter $\delta$ adds randomness to the system, causing frequent, unpredictable changes in entanglement, with both positive and negative values of $\delta$ leading to similar effects. These fluctuations indicate continuous cycles of generating and destroying entanglement, suggesting a strong exchange of information between the system and its surrounding fields. Overall, the results show that the gravcat state is highly dynamic, with its entanglement deeply affected by environmental conditions and system interactions.}

\subsection{General local decaying field}\label{GLDF}
Let us consider a general classical decaying field of the type $\mu \exp[-\chi t]$ and replace it with the coupling constant $\lambda$ given in Eq. \eqref{rho-t}, i.e. 
\begin{equation}
\lambda:=\mu \exp[-\chi t]\label{decaying field}
\end{equation}
where the parameter $\mu$ remains the coupling strength of the decaying field with the system while $\chi$ regulates the decay rate and $t$ is the time of the coupling \cite{rahmanAEJ}. 

{ This classical decaying field application controls the decay rate of the interaction between the system and the environment. This addresses the physical scenario where the interaction strength between the quantum system (in our case gravcat system) and its environment diminishes with time, which can arise in many physical settings, especially in systems subject to time-dependent noise or external fields that decay in strength. This decaying interaction is to mimic physical systems whereby the interaction with the environment decreases due to some factors such as dissipation, attenuation, scattering, absorption or distance from a source of the interaction. As such, a decaying interaction is typical of conditions that are perturbed in practice with reduced amplitudes of external control of noise field influences the degree of entanglement of the coupled system. In this regard, the authors of Ref. \cite{x1} demonstrated fidelity variations under the influence of decoherence in terms of leakage from electrical and cavity fields. Another example where the coupling between the qubits and environment leads to a change in the degree of entropy is discussed in Refs. \cite{x2, x3}. More specifically, decoherence between sub-quantum systems has been shown to induce against varying the coupling strength \cite{x3a, x3b}. Besides, there are various methodologies present in quantum optics and solid-state systems where one can experimentally vary the coupling strengths of classical or other fields, for example, modulating the phase of the field \cite{x4}, processes of adiabatic switching on/off the interaction of a two-level atom (at rest) with a quantized mode \cite{x5}, etc. Therefore, variations in coupling strength can significantly influence the degree of decoherence, leading to a degradation of relative entanglement and other quantum correlations. It should be noted that, in our context, the term ``decay" serves as an indirect indicator of entanglement degradation. This degradation occurs due to leakage or modulation effects of the classical field interacting with the system.}

Again, we employ the EW used in the previous section to evaluate entanglement decay caused by the current classical decaying field. The result obtained for the EW is the same as that given in Eq. \eqref{EW-reults}, but with
\begin{equation}
\Upsilon_4=\gamma ^2 \cos (4 \delta \mu e^{-\chi t}  t) \sinh ^2\left(\mathcal{K}/T\right).
\end{equation}

In the following, we discuss the dynamics of entanglement in the gravcat state when subjected to a classical decaying field.
\begin{figure*}[!t]
	\begin{center}
		\includegraphics[width=0.32\textwidth, height=145px]{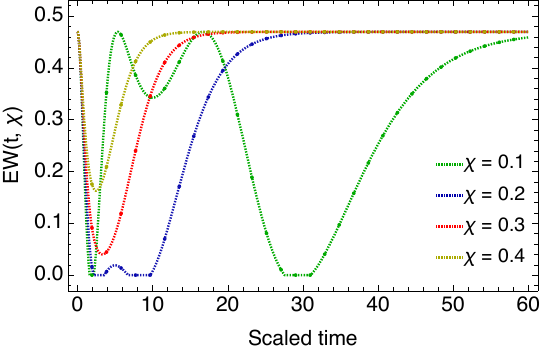}
		\put(-145,150){($ a $)}\quad
		\includegraphics[width=0.32\textwidth, height=145px]{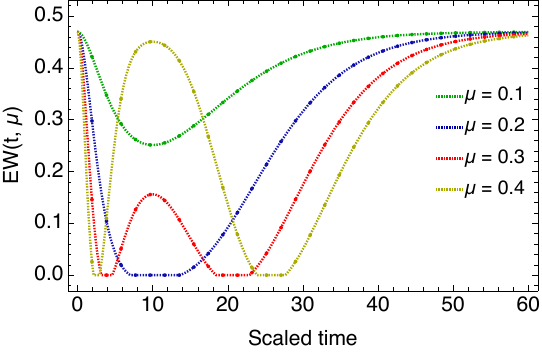}
		\put(-145,150){($ b $)}\quad
		\includegraphics[width=0.32\textwidth, height=145px]{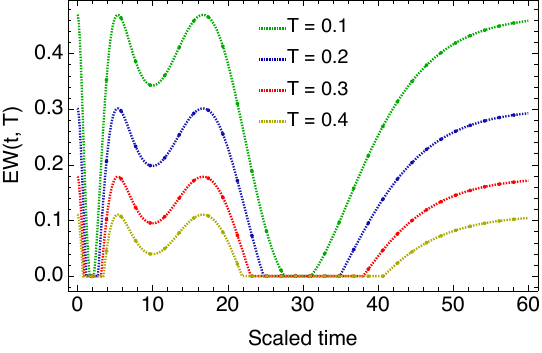}
		\put(-145,150){($ c $)}\\ [0.2cm]
		\includegraphics[width=0.32\textwidth, height=145px]{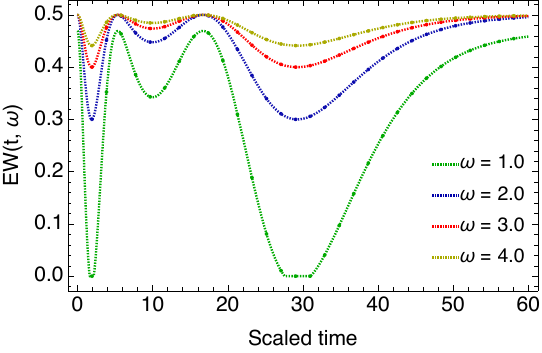}
		\put(-145,150){($ d $)}\quad
		\includegraphics[width=0.32\textwidth, height=145px]{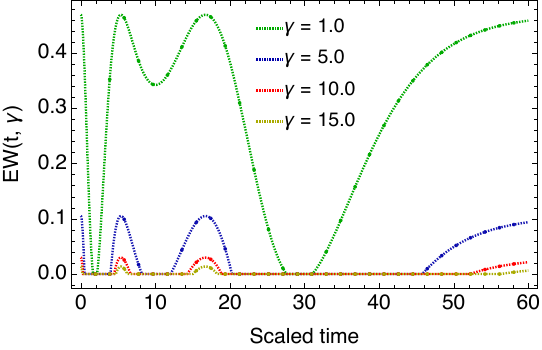}
		\put(-145,150){($ e $)}\quad 	
		\includegraphics[width=0.32\textwidth, height=145px]{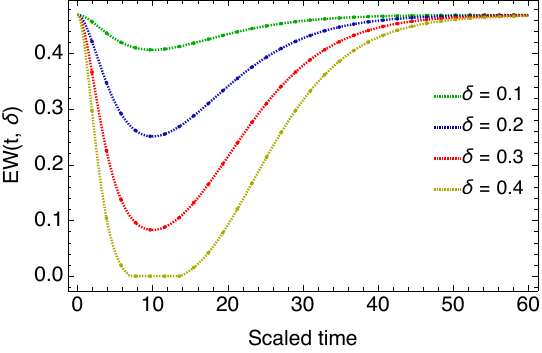}
		\put(-145,150){($ f $)}		
		\end{center}
\caption{Dynamics of EW for the two-qubit gravcat state prepared in the state $\rho(t, T)$ given in Eq. \eqref{rho-t} influenced by a classical decaying field when (a) $\delta=\gamma=\omega=1,~\mu=0.5,~T =0.1$, (b) $\delta=\gamma=\omega=1,~ \chi=0.1,~T =0.1$, (c) $\delta=\gamma=\omega=1,~\chi =0.1,~\mu=0.5$, (d) $\delta=\gamma=1,~\chi =0.1,~\mu=0.5, ~T=0.1$, (e) $\delta=\omega=1,~\chi =0.1,~\mu=0.5,~T=0.1$, and (f) $\gamma=\omega=1,~\chi =0.1,~\mu=0.5,~T=0.1$.}\label{fig7}
\end{figure*}
In Fig. \ref{fig7}(a), the EW is primarily characterized by the decay parameter $\chi$. Initially, the state has the equal amount of entanglement even for the different values of $\chi$. However, with time, the decay becomes variant for the different values of $\chi$. It is interesting to notice that for the lower $\chi$ values, entanglement remains vanished for a longer time after the drop compared to that observed for the higher $\chi$ values. As seen for the higher $\chi$ values, entanglement faces a sudden death for a shorter time.  Notably, $\chi=0.3$ is the most favorable situation for higher entanglement preservation.
\par
In Fig. \ref{fig7}(b), the impact of coupling strength between the decaying field and system is studied. One can easily demonstrate the weak and strong coupling strengths have no impact on the initial value of EW. However, after the onset, the entanglement suffers a greater decay in the strong coupling regimes.
\par
The influence of temperature on the EW in the gravcat state is evaluated in Fig. \ref{fig7}(c). As expected, the initial entanglement values in the state remain higher for the lower temperature values. But, the EW function remains more suppressed as the temperature rises, negatively affecting the initial level of entanglement.
\par
In Fig. \ref{fig7}(d), the impact of the difference between the excited and ground state is studied on the EW in the gravcat state. Initially, the parameter $\omega$ has a direct impact on the degree of entanglement and for the increasing values of $\omega$, entanglement in the state increases and vice versa. With time, for the higher $\omega$ values, the decay is least and the state remains highly entangled. For the least $\omega$ value (e.g. $\omega=1.0$), the state enters the disentanglement regime for a longer interval of time.
\par
In Fig. \ref{fig7}(e), the impact of gravitational interaction strength $\gamma$ on the EW of the gravcat state is investigated. Initially, as well as for the later interval of time, the state is detected to be in the stronger entanglement regime. In comparison, the state becomes separable when $\gamma$ is raised to higher strengths.
 \par
Finally,  the influence of the stochastic parameter $\delta$ on the dynamics of gravcats entanglement is studied in Fig. \ref{fig7}(f). It is interesting to note that for the lower $\delta$ values, the entanglement in the gravcat state suffers a lesser drop as compared to the higher strengths of $\delta$. \par

It is worth mentioning that at the given final notes of time, the initial level of entanglement is regained. Hence, the current classical field and the considered configuration are good choices for controlling quantum correlations in practical quantum information processing.

\subsection{Power-law noisy dephasing channel}\label{power-law}
Here, we give the details of the application of the power-law (PL) noise \cite{38} on the dynamics of the thermal density matrix. The phase of the single-qubit system can be denoted by $ \phi_n (t)=- \lambda \int_{0}^{t}\delta_n(s) ds$ \cite{36}. Let us define the $\beta$-function, which incorporates the phase factor of the noise with the phase of the system and can be written as \cite{39}
\begin{equation}
\beta(t)=\int_0^t \int_0^t K(s-s^{\prime})ds ds^\prime. \label{Beta function}
\end{equation}
The related auto-correlation function of the PL noise by considering the noise parameter $\alpha$ can be written as \cite{rahman5342022}
\begin{equation}
K(s-s^{\prime},\zeta,\theta,\alpha)=\frac{\theta \zeta (\alpha-1)}{2[1+\theta \vert s-s^{\prime}\vert]^2}.\label{AC of PL}
\end{equation}
By assuming the dimensionless terms as $g=\theta/\zeta$ as well as $\tau=\zeta t$  and inserting the auto-correlation function given in Eq.~\eqref{AC of PL} into Eq. \eqref{Beta function}, the $\beta$-function of the PL noise takes the form \cite{37}
\begin{equation}
\beta_{PL}(\tau)=\frac{g \tau(\alpha -2)-1+(1+g \tau)^{2-\alpha }}{(\alpha -2)g}.\label{beta function of PL}
\end{equation}
The Gaussian function with zero mean has the form $\langle e^{\pm i m\phi_{n}(t)}\rangle = e^{\eta_{n}(\tau)}$ where $\eta_{n}(\tau)=-\frac{1}{2}m^2\beta_{PL}(\tau)$ with $n, m \in \mathbb{N}$, which can be used to employ a Gaussian noise phase over the joint phase of the system and classical field \cite{36, 37}. To find the noisy effects, we take an average of the thermal density matrix of gravcats $\rho(0, T)$ given in Eq. \eqref{eq:rho-mat} over the PL noise phase as

\begin{equation}
\rho(t, T):=\left\langle U(t)\rho(0, T) U^{\dagger}(t)\right\rangle_{{\eta_n(\tau)}},
\label{final density matrix2}
\end{equation}
where $U(t)$ is defined in Eq. \eqref{ut}.
Hence, the explicit form of the final density matrix can be written as
\begin{equation}
\rho(t, T)=\left[
\begin{array}{cccc}
 \rho_{11} & 0 & 0 &  e^{-8 \beta_{PL}(\tau)} \rho_{14}\\[0.2cm]
 0 & \rho_{22} & \rho_{23} & 0 \\[0.2cm]
 0 & \rho_{23}^* & \rho_{22} & 0 \\[0.2cm]
 e^{-8 \beta_{PL}(\tau)} \rho_{14}^* & 0 & 0 & \rho_{44} \\
\end{array}
\right].\label{final density matrix3}
\end{equation}

\subsubsection{Power-law noise assisted with decaying field}
In Sec. \ref{GLDF}, we focused only on the decay effects caused by the general local field. Here, we intend to provide the combined effects of PL and local decaying field on the initial thermal state. In this regard, we first impose the PL noise on the dynamics of the system \eqref{final density matrix3} and additionally assist with classical decaying field \eqref{decaying field}. Next, we use the EW measure given in Eq.  \eqref{ewww} to provide a quick comparison between the impact of solely employed classical decaying field and PL noise added with the decaying field.

\begin{figure*}[!t]
	\begin{center}
		\includegraphics[width=0.32\textwidth, height=145px]{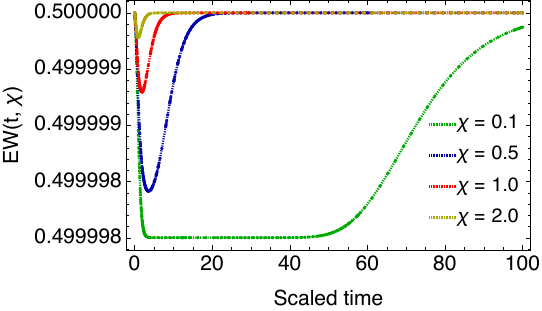}
		\put(-145,150){($ a $)}\quad
		\includegraphics[width=0.32\textwidth, height=145px]{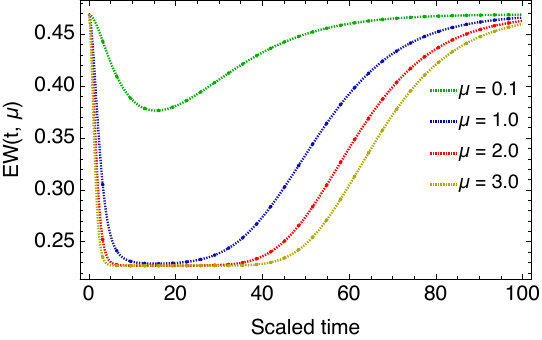}
		\put(-145,150){($ b $)}\quad
		\includegraphics[width=0.32\textwidth, height=145px]{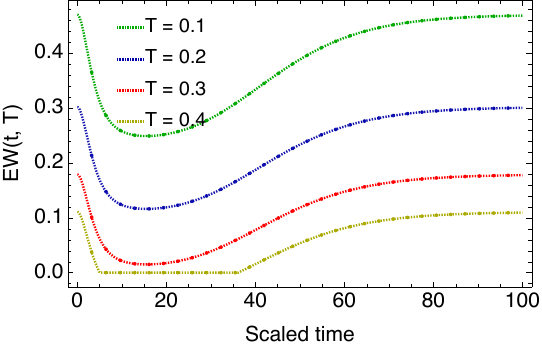}
		\put(-145,150){($ c $)}\\ [0.2cm]
		\includegraphics[width=0.32\textwidth, height=145px]{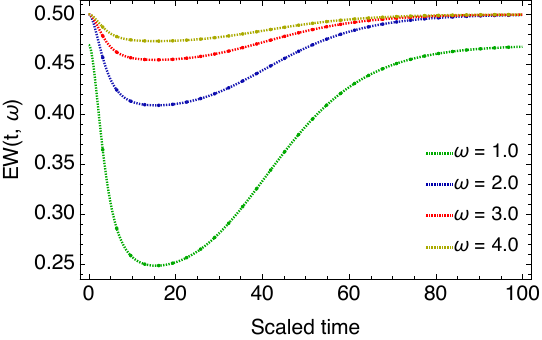}
		\put(-145,150){($ d $)}\quad
		\includegraphics[width=0.32\textwidth, height=145px]{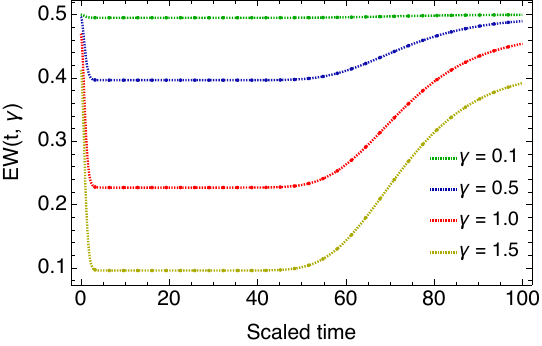}
		\put(-145,150){($ e $)}\quad 	
		\includegraphics[width=0.32\textwidth, height=145px]{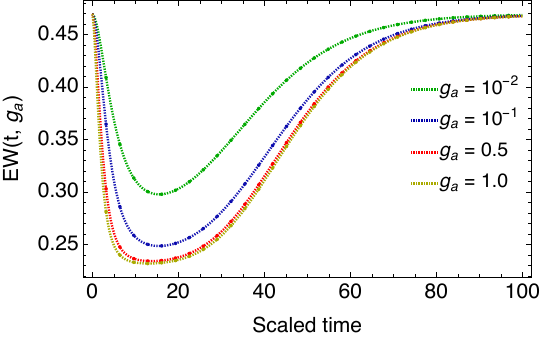}
		\put(-145,150){($ f $)}		
		\end{center}
\caption{ Dynamics of EW for the two-qubit gravcat state prepared in the state $\rho(t, T)$ given in Eq. \eqref{rho-t} influenced by PL noise assisted with classical decaying field when (a) $~g_a=g_b=T=\gamma=10^{-1}$, $\omega=\mu=5$, (b) $g_a=g_b=\chi=T=\gamma=10^{-1},~\omega=5$, (c) $g_a=g_b=\chi=\gamma=10^{-1}$, $\mu=\omega=5$, (d) $g_a=g_b=T=\chi=\gamma=10^{-1}$, $\mu=5$, (e) $g_a=g_b=\chi=T=10^{-1}$, $\omega=\mu=5$, and (f) $g_b=\chi=T=\gamma=10^{-1},~\mu=\omega=5$.}\label{figxx}
\end{figure*}
Figure \ref{figxx} shows the time evolution of  EW for a two-qubit gravcat state subjected to PL noise assisted with the classical decaying field. Here, we illustrate how the inclusive parameters influence the entanglement over the scaled time, revealing distinct behaviors that indicate the strength and character of dephasing effects in various perturbative regions. { For the perturbative regime ($\gamma \ll \{\omega, \lambda\}$), the dephasing effects are small, which leads to a slow decay of entanglement and high values of the EW during the dynamics.} In Fig. \ref{figxx}(a, d), the higher values of $\chi$ and $\omega$ help to reduce decay with time. In Fig. \ref{figxx}(b, c, e, f), the relative smaller values of $\mu,~T,~\gamma,~g_a/g_b$ are resourceful for quantum correlations preservation.  In all cases, the entanglement is high at the beginning; it is reduced slightly and then it settles to a slightly lower value compared to the initial entanglement limit. It is important to note that when $\gamma \approx \{\omega, \lambda\}$ or larger, as in Fig. \ref{figxx}(e), the dephasing effects are more significant. { This higher decay rate results from the system being more susceptible to noise at large $\gamma$ values and in a non-perturbative regime ($\gamma \gg \{\omega, \lambda\}$).  The obtained results suggest that entanglement is better protected in the perturbative regime, and is almost negligible. In comparison, within a non-perturbative regime, the noise-induced dephasing fastens the rate of entanglement decay [e.g., see Fig. \ref{fig7}(e)]. This means that the choice of perturbative/non-perturbative regimes on the current system dynamics greatly affects the decay rate, even compared to the role of noise.}

\subsubsection{Quantum correlations estimators}

\textbf{Steerability}:
Let us assume an arbitrary two-qubit density matrix having the Bloch decomposition, which has the form \cite{40}
\begin{align}\label{xxx}
&\rho_{ab}=\nonumber\\
&\frac{1}{4} \bigg[ \mathbb{I}_a \otimes \mathbb{I}_b+ \vec{\mathcal{A}}. \sigma_a \otimes \mathbb{I}_b+ \mathbb{I}_a \otimes \vec{\mathcal{B}}.\sigma_b + \sum_{i,j=1}^{3} \mathcal{C}_{ij}\ \sigma^i_a \otimes \sigma^j_b\bigg],
\end{align}
where $\sigma_{a,b} $ denote Pauli operators, $  \vec{\mathcal{A}} $ and $ \vec{\mathcal{B}} $ are the Bloch vectors with $ \mathcal{A}_i={\rm Tr}[\rho_{ab}.(\sigma^i \otimes \mathbb{I}_b )] $ and $ \mathcal{B}_i={\rm Tr}[\rho_{ab}.(\mathbb{I}_a \otimes \sigma^i )] $, while $ \mathcal{C}_{ij}= {\rm Tr}[\rho_{ab} \sigma^i \otimes \sigma^j ]  $. In the computational basis $ \{|00\rangle, |01\rangle,|10\rangle, |11\rangle \} $, the two-qubit $X$--shaped density matrix $ \rho_{ab} $ can be written as
 \begin{equation}\label{fpr}
 \rho_{ab}=\left[
	\begin{array}{cccc}
		\varrho_{11} & 0 & 0 & \varrho_{14} \\[0.2cm]
		0 & \varrho_{22} & \varrho_{23} & 0 \\[0.2cm]
		0 & \varrho_{23}^* & \varrho_{33} & 0 \\[0.2cm]
		\varrho_{14}^* & 0 & 0 & \varrho_{44}\\ [0.2cm]
	\end{array}
	\right],
\end{equation}
where $ \rho_{ij}^* $ are the Hermitian conjugates of $ \rho_{ij} $. Besides
\begin{align*}
\mathcal{A}_1=&\mathcal{A}_2=0, \qquad
\mathcal{A}_3=\varrho_{11}+\varrho_{22}-\varrho_{33}-\varrho_{44},  \\
\mathcal{B}_1=&\mathcal{B}_2=0, \qquad
\mathcal{B}_3=\varrho_{11}-\varrho_{22}+\varrho_{33}-\varrho_{44}, \\
\mathcal{C}_{11}= &2 {\rm Re}[\varrho_{23}+\varrho_{14}], \qquad
\mathcal{C}_{22}= 2 {\rm Re}[\varrho_{23}-\varrho_{14}],\\
\mathcal{C}_{33}=&\varrho_{11}-\varrho_{22}-\varrho_{33}+\varrho_{44}, \qquad
\mathcal{C}_{ij}=0\ \text{if} \ i\neq j.
\end{align*}	

In 2013, Schneeloch {\it et al.} \cite{41} defined the steering inequality for the two-qubit state using the Shannon entropy as well as Pauli spin matrices as discrete observables. It has the form
\begin{equation} \label{e1}
I_{ab}:=\mathcal{H}(\sigma_x^b|\sigma_x^a)+\mathcal{H}(\sigma_y^b|\sigma_y^a)+\mathcal{H}(\sigma_z^b|\sigma_z^a)\geq 2,
\end{equation}
where $\mathcal{H}(R^b|R^a)=\mathcal{H}(\rho^{ab})- \mathcal{H}(\rho^{a})$ represents the conditional Shannon entropy. For the density matrix of the two-qubit state, we have $I_{ab}\in \{0,~6\} $. Note that the two-qubit state $ \rho_{ab} $ will be non-steerable for $ {I}_{ab}\leq 2 $. Now, by employing Eq. (\ref{e1}) for the $X $-shaped density matrix, the two-qubit steering inequality takes the form
\begin{equation}
	\begin{split}
		I_{ab}= &\sum_{i=1,2}\left([1-\mathcal{C}_{ii}] \log_2 [1-\mathcal{C}_{ii}]+ [1+\mathcal{C}_{ii}] \log_2 [1+\mathcal{C}_{ii}]\right)\\
  &- \mathcal{X}_1+\frac{1}{2}[\mathcal{X}_2+ \mathcal{X}_3+ \mathcal{X}_4+\mathcal{X}_5 \big],
	\end{split}
\end{equation}
where
\begin{align*}
\mathcal{X}_1=&(1+\mathcal{A}_3) \log_2 (1+\mathcal{A}_3)+(1-\mathcal{A}_3) \log_2 (1-\mathcal{A}_3),\\
\mathcal{X}_2=&(1+\mathcal{C}_{33}+\mathcal{A}_3+\mathcal{B}_3) \log_2 (1+\mathcal{C}_{33}+\mathcal{A}_3+\mathcal{B}_3),\\
\mathcal{X}_3=&(1-\mathcal{C}_{33}-  \mathcal{A}_3+\mathcal{B}_3) \log_2 (1-\mathcal{C}_{33}-\mathcal{A}_3+\mathcal{B}_3),\\
\mathcal{X}_4=&(1+\mathcal{C}_{33}-\mathcal{A}_3-\mathcal{B}_3) \log_2 (1+\mathcal{C}_{33}- \mathcal{A}_3-\mathcal{B}_3),\\
\mathcal{X}_5=&(1-\mathcal{C}_{33}+\mathcal{A}_3-\mathcal{B}_3) \log_2 (1-\mathcal{C}_{33}+\mathcal{A}_3-\mathcal{B}_3).
\end{align*}

Hence, the one-way steerability from qubit $a \rightarrow b$ can be done by introducing the expression
\begin{equation}
{\rm ST}:=\max\bigg[0,~\frac{I_{ab}-2}{{I}_{max}-2}\bigg],\label{ST}
\end{equation}
where ${I}_{max}=6$ and the steerability for the two-qubit states has the range $0\leq {\rm ST}\leq1$.

\textbf{Bell non-locality}:
The distinctive feature of quantum systems that results from the violation of Bell-like inequalities is Bell non-locality. Additionally, no local hidden variable approach can adequately capture Bell non-locality. The Clauser--Horne--Shimony--Holt (CHSH) inequality, which can be employed in the proof of Bell's theorem, is of special relevance since it is simple to compute for a two-qubit system and has the form \cite{ 42, ML1}
\begin{align}
B(\rho_{ab}, \xi)= \vert {\rm Tr}\left(\rho_{ab} B_{\textmd{CHSH}}\right) \vert \leq 2, \label{B_{SHCH}}
\end{align}
where $B_{\textmd{CHSH}}$ denotes a Bell-CHSH operator acting on ${\cal H}_a \otimes {\cal H}_b$, and $\xi$ is the measurement configuration for two users, Alice and Bob, which is obtained classically via local hidden-variable model. However, it must satisfy the inequality condition for the two-qubit density matrix $\rho_{ab}$. Therefore, the Bell non-locality measure for $\rho_{ab}$ can be written as \cite{40}
\begin{equation}
{\rm BN}:=\max_{\xi}\left[B(\rho_{ab}, \xi)\right],
\end{equation}
and finally, one can write the normalized Bell non-locality function as
\begin{equation}
{\rm BN}:=\max\Bigg[ 0, \frac{B_{\textmd{CHSH}}-2}{\mathcal{B}_{\max}-2}\Bigg].\label{BN}
\end{equation}

Note that for the given two-qubit $X$--shaped density matrix \eqref{fpr}, we have
$B_{\textmd{CHSH}}=2 \max \left\{b_1 \mathcal{B}_{max}, b_2\right\}$
where
$b_1=\sqrt{|\varrho_{14}|^2+|\varrho_{23}|^2}$ and $b_2=\sqrt{4(|\varrho_{14}|+|\varrho_{23}|)^2+\mathcal{C}_{33}^2}$  with $\mathcal{B}_{max}=2\sqrt{2}$. Hence, the normalized Bell non-locality function \eqref{BN} has the range $0\leq {\rm BN}\leq1$ since $B_{\textmd{CHSH}}\leq \mathcal{B}_{max}$.

\textbf{Concurrence}:
Concurrence is a measure of entanglement for quantum systems composed of two qubits \cite{43}. It quantifies the degree of entanglement between the two qubits in a quantum state.
For a quantum state described by a two-qubit $X$--shaped density matrix $\rho_{ab}$  \eqref{fpr}, the concurrence is formulated as follows

\begin{equation}
\operatorname{CN}:=2\max \left[0,~ |\varrho_{23}|-\sqrt{\varrho_{11}\varrho_{44}},~|\varrho_{14}|-\sqrt{\varrho_{22}\varrho_{33}} \right].\label{CN}
\end{equation}

Notice, the concurrence ranges from 0 to 1, where zero indicates no entanglement, and 1 indicates maximum entanglement.

\textbf{Purity}:
Purity is an important concept in quantum information theory and quantum computing. It is used in various contexts, such as quantifying the amount of entanglement in a quantum system, characterizing the quality of quantum states in quantum algorithms, and analyzing the effects of noise and decoherence in quantum computations. A high purity in quantum systems is desirable, as it indicates a system with reduced decoherence and greater coherence, which is essential for the reliable implementation of quantum algorithms and quantum information processing tasks.
Mathematically, the purity of a quantum state with density operator $\rho_{ab}$ is given by
\begin{equation}
{\rm PR} :={\rm Tr}\left[\rho_{ab}^{2} \right]. \label{PR}
\end{equation}

A purity of 1 indicates that the quantum system is in a pure state, while a purity of less than one shows that the system is in a mixed state with some degree of uncertainty or entropy.

\begin{figure*}[!t]
	\begin{center}
		\includegraphics[width=0.47\textwidth, height=145px]{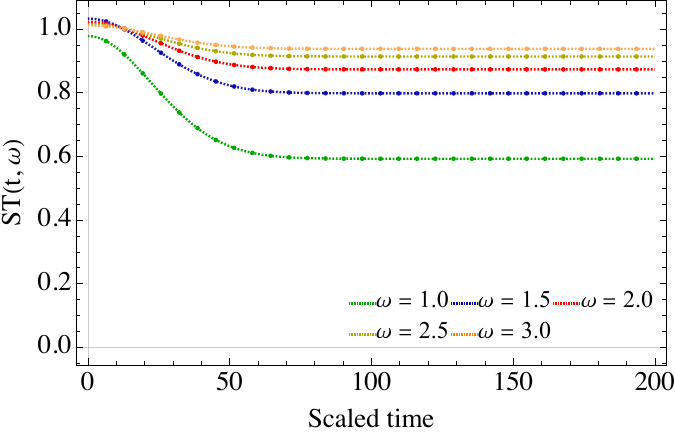}
		\put(-200,148){($ a $)}\quad
		\includegraphics[width=0.47\textwidth, height=145px]{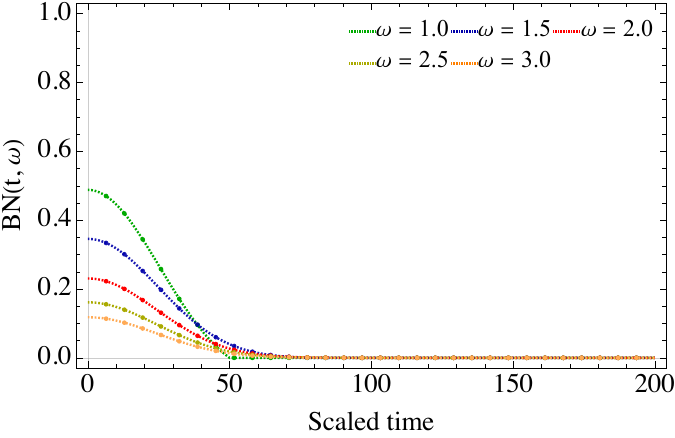}		\put(-200,148){($ b $)}\\
		\includegraphics[width=0.47\textwidth, height=145px]{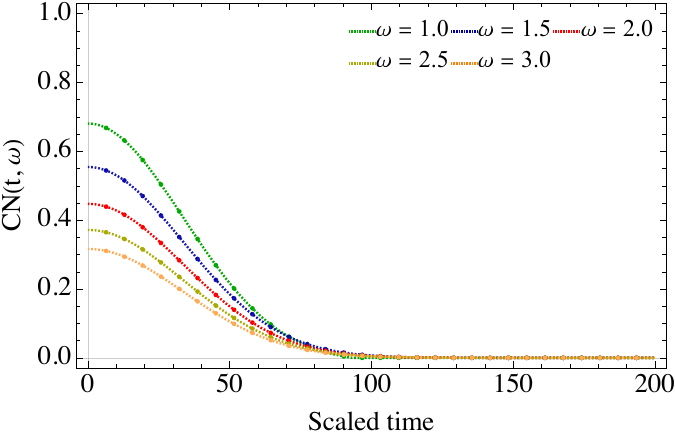}
		\put(-200,148){($ c $)}\quad
		\includegraphics[width=0.47\textwidth, height=145px]{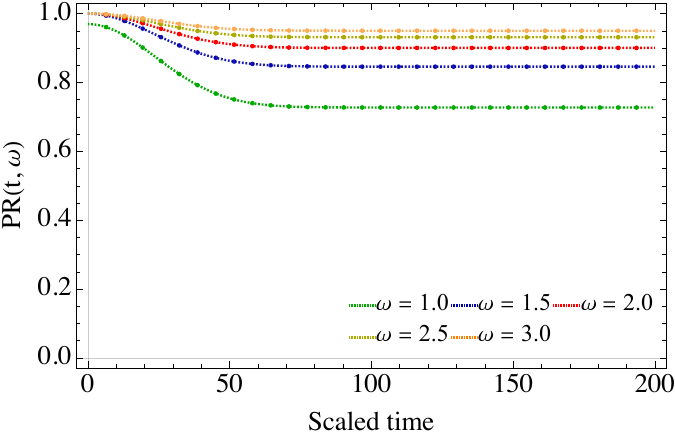}
		\put(-200,148){($ d $)}\quad
		\end{center}
\caption{Steerability (a), Bell non-locality (b), concurrence (c) and purity (d) in the thermal two-qubit gravcat state when influenced by a PL noise versus scaled time and fixed values of $\omega$  for  $\gamma =1$,  $\alpha =2.1$, $g=10^{-4}$, and $T= 0.1$.}\label{fig1}
\end{figure*}

Now, we can study the dynamics of quantum correlations in the thermal two-qubit gravcat state when influenced by a PL noise using the final density matrix of the system given in Eq. \eqref{final density matrix3} and the formula for ST \eqref{ST}, BN \eqref{BN}, CN \eqref{CN}, and PR \eqref{PR}.

In Fig. \ref{fig1}, the impact of various fixed distance values ($\omega$) between the ground state $\ket{g}$ and excited state $\ket{e}$  on the dynamics of the two-qubit gravcat state is presented. In particular, the degree of steerability, non-locality, entanglement, and purity are aimed to be investigated when the thermal gravcat state is influenced by a PL noisy channel. At the onset, the gravcat state exhibits different values of steerability, non-locality, entanglement, and purity depending upon the parameter setting of $\omega$. In the case of ST and PR, the state shows enhancement in the initial level of steerability and purity for the increasing $\omega$ values. However,  CN and BN  show that for $\omega=1.0$, the state exhibits higher entanglement and non-locality values. This means the different nature of the two-qubit quantum correlations.
Besides, as the gravcats correlations are allowed to evolve with time, the ST, BN, CN, and PR functions show a decline, suggesting the emergence of the dephasing effects in the thermal two-qubit gravcat state. However, the steerability and purity of the gravcats finally stabilize and evolve at a constant rate without showing any further loss.

It is noticeable that with the increasing distance between the excited and ground states, the final preserved values of steerability, non-locality, entanglement, and purity are enhanced. In comparison, BN remains extremely fragile to the dephasing effects of the  PL noise and vanishes for all $\omega$ values. Under the influence of the PL noise, the quantum correlations (BN and CN) reveal deaths without rebirth. This means that the information is lost when the PL noise is employed. The convenient reason for this is the no information backflow between the system and the coupled channel. Therefore, the PL noise shows no resources for the conversion of classical product states into quantum entangled states. The degree of steerability, and purity while non-locality, and entanglement in the state have provided qualitatively agreeing results, hence, suggesting a connection between them.

\begin{figure*}[!t]
	\begin{center}
		\includegraphics[width=0.47\textwidth, height=145px]{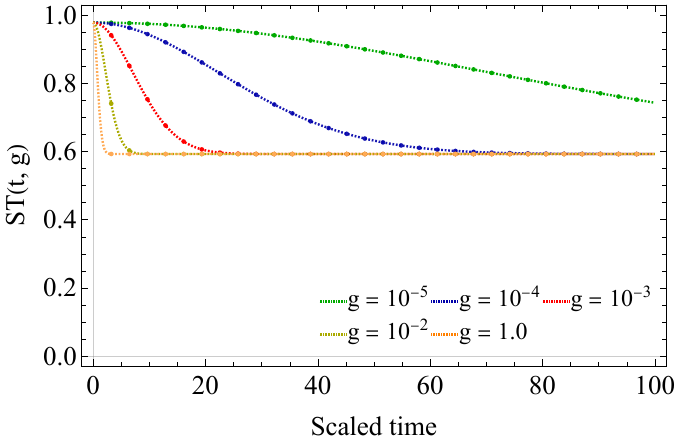}
		\put(-200,148){($ a $)}\quad
		\includegraphics[width=0.47\textwidth, height=145px]{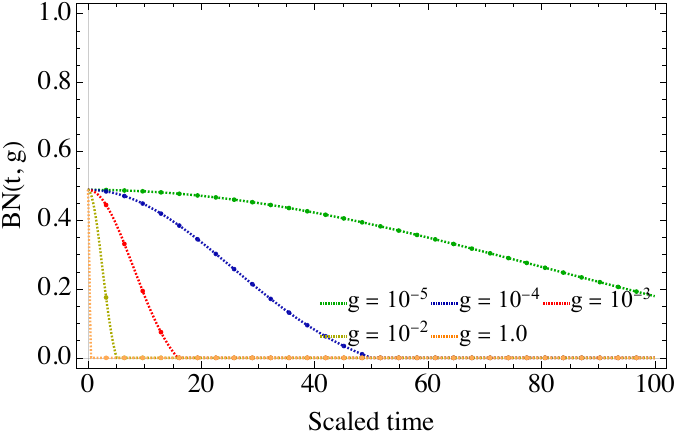}
		\put(-200,148){($ b $)}\\
		\includegraphics[width=0.47\textwidth, height=145px]{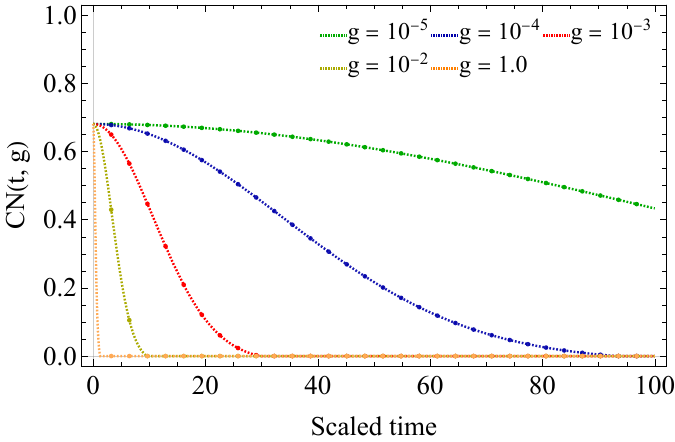}
		\put(-200,148){($ c $)}\quad
		\includegraphics[width=0.47\textwidth, height=145px]{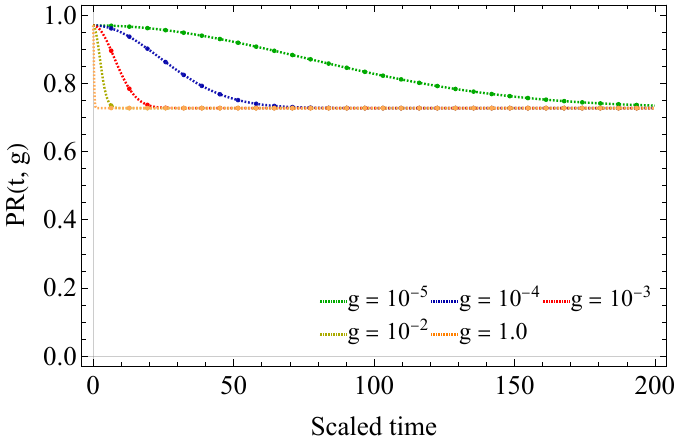}
		\put(-200,148){($ d $)}\quad
		\end{center}
\caption{Steerability (a), Bell non-locality (b), concurrence (c) and purity (d) are shown as in Fig. \ref{fig1}  but for fixed values of $g$ with  $\omega =1$.}\label{fig2}
\end{figure*}

Figure \ref{fig2} illustrates the dynamics of ST, BN, CN, and PR versus the fixed noise parameter $g$ values in a two-qubit gravcat state when exposed to the dephasing effects. {  At the beginning, for different values of $g$, all measures (ST, BN, CN, PR) are seen to remain constant. This suggests that the initial degree of steerability, non-locality, entanglement, and purity in the gravcat state is unaffected by the classical field noise strength parameter $g$. This stability physically indicates that the system's initial state is sufficiently isolated or robust against the initial application of the dephasing noise. As time progresses, the gravcat correlations decrease exponentially. This reveals a continuous loss of quantum coherence and correlations in the system, caused by the classical dephasing noise. The parameter $g$, which regulates the noise's strength, plays a crucial role in the rate at which these correlations decay. The more significant the noise parameter strength ($g$), the faster the loss of quantum properties occurs. This exponential decay is a hallmark of dephasing processes, where phase relationships between the quantum states are randomly disrupted over time, leading to a reduction in the overall quantum correlations. 

Eventually, the mentioned functions (ST, BN, CN, PR) reach saturation points where they stabilize at constant levels. Physically, this can be interpreted as the system reaching a state of equilibrium with the thermal environment. The saturation indicates that no further exchange of information or quantum attributes is taking place between the gravcat state and the external noise field. The exponential decrease of quantum correlations suggests no exchange of information. This implies that the dephasing noise irreversibly destroys the quantum correlations, with no possibility for the field to restore or generate new non-local correlations in the gravcat state. This behavior demonstrates the fragility of quantum states like the gravcat under classical noise, which is a fundamental limitation in maintaining long-term quantum coherence and entanglement in such systems. 

Moreover, the exponential decay and saturation of quantum correlations in the presence of classical dephasing noise (as seen in the current case) directly relate to practical challenges in quantum computing and information. Specifically, noise-induced decoherence leads to the irreversible loss of entanglement, steerability, and non-locality, critical for tasks like quantum communication, error correction, and maintaining coherence in qubits, thus emphasizing the need for robust noise-resisting methodologies to be employed.}

\begin{figure*}[!t]
	\begin{center}
		\includegraphics[width=0.47\textwidth, height=145px]{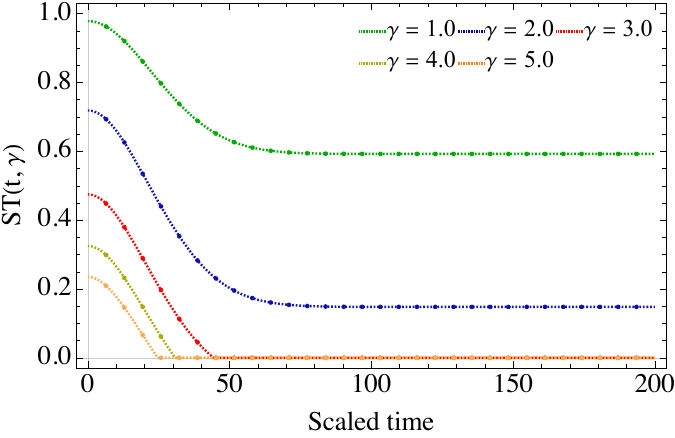}
		\put(-200,148){($ a $)}\quad
		\includegraphics[width=0.47\textwidth, height=145px]{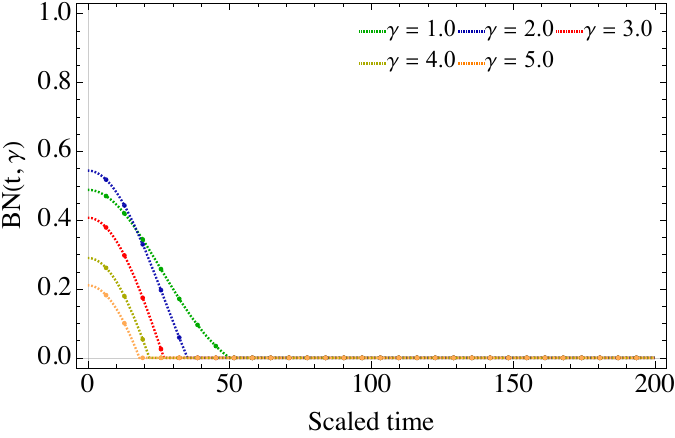}
		\put(-200,148){($ b $)}\\
		\includegraphics[width=0.47\textwidth, height=145px]{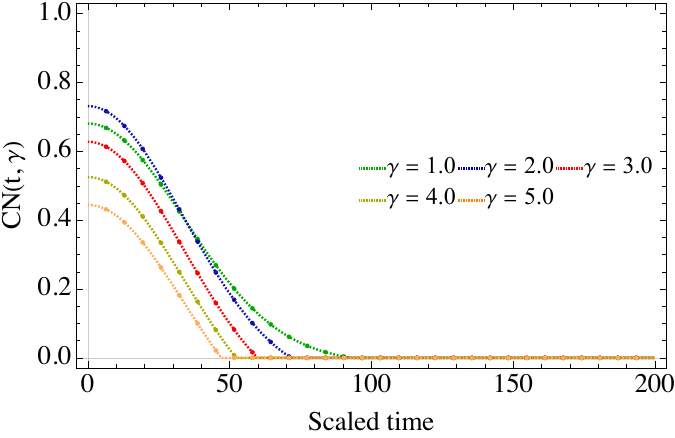}
		\put(-200,148){($ c $)}\quad
		\includegraphics[width=0.47\textwidth, height=145px]{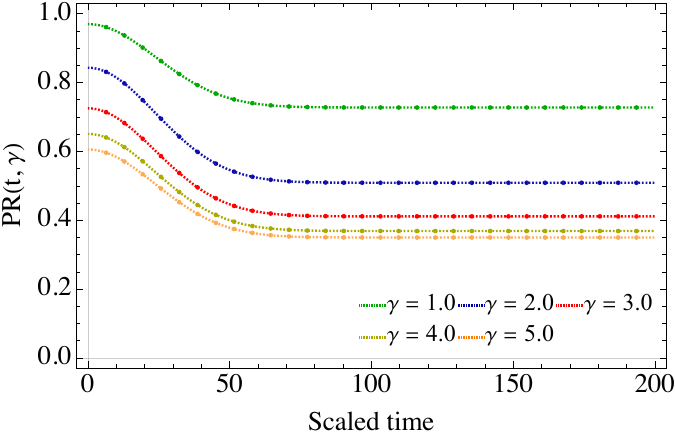}
		\put(-200,148){($ d $)}\quad
		\end{center}
\caption{Steerability (a), Bell non-locality (b), concurrence (c) and purity (d) are shown as in Fig. \ref{fig1}  but for fixed values of $\gamma$ with  $\omega =1$.} \label{fig3}
\end{figure*}

Besides, in agreement with Fig. \ref{fig1}, the steerability and pureness in the state remain preserved. However,  non-locality and entanglement in the gravcat state show less tolerance against the dephasing effects of the PL noise and become lost, even for the lower PL noise parameter strengths. This is important to note because the PL noise parameter has an extreme impact on the declining time of the gravcat correlations until reaching an asymptotic final saturation level. For example, for the lower values of $g$, the declining rates of steerability, non-locality, entanglement, and pureness in the state become slower. Unlike this, the higher $g$ values greatly accelerate the decrease in the two-qubit gravcat correlations. Hence, the dephasing effects of the PL noise are highly enhanced by increasing $g$. Notice, the PL noise parameter $g$ shows an inverse effect on the quantum correlations compared to $\omega$. As the first one degrades and the latter one enhances quantum correlations.

Fig. \ref{fig3} discloses the time evolution of the steerability, non-locality, entanglement, and pureness of the gravcat state when governed by the dephasing effects of PL noise.
The initial quantum correlations in the state reduce with time and finally saturate at final levels. However, the absolute decay in the two-qubit steerability, non-locality, entanglement, and purity remains variable against the different values of $\gamma$. For example, in all the cases, the non-classical correlations functions remain more robust and face smaller decay for the weaker gravitational strength. However, as the two qubits gravitate more toward each other, the resourcefulness of the state seems lost. Interestingly, the steerability between the two qubits is affected the most compared to non-locality, entanglement, and purity.
This suggests that individual parameter gauging and characterization are vital tasks in different situations to improve quantum correlations. Finally, in connection with Figs. \ref{fig1} and \ref{fig2}, the decay is fully exponential, either with a single death point or remains partially preserved for a longer interval of time.

\begin{figure*}[!t]
	\begin{center}
		\includegraphics[width=0.47\textwidth, height=145px]{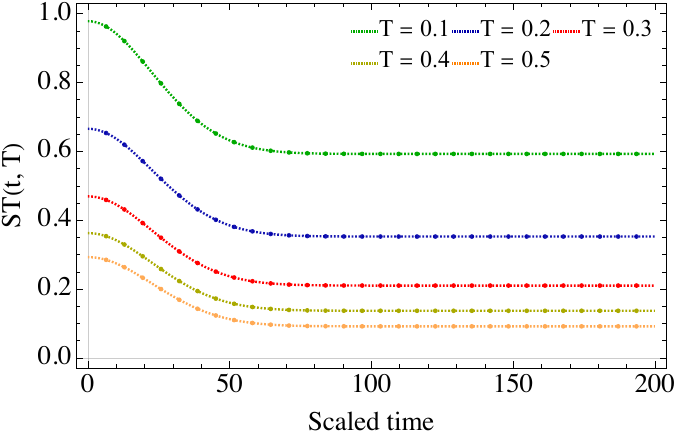}
		\put(-200,148){($ a $)}\quad
		\includegraphics[width=0.47\textwidth, height=145px]{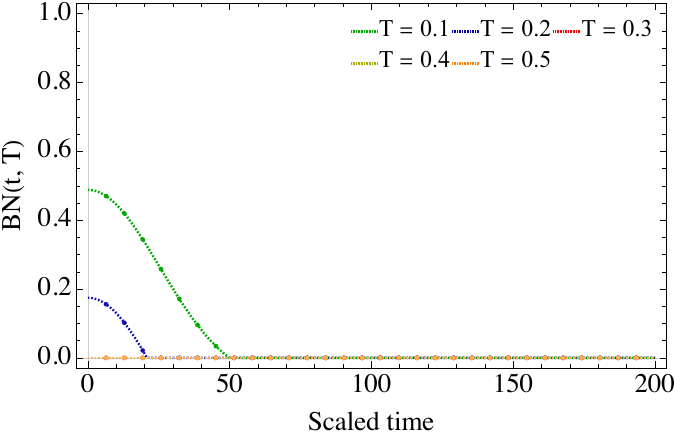}
		\put(-200,148){($ b $)}\\
		\includegraphics[width=0.47\textwidth, height=145px]{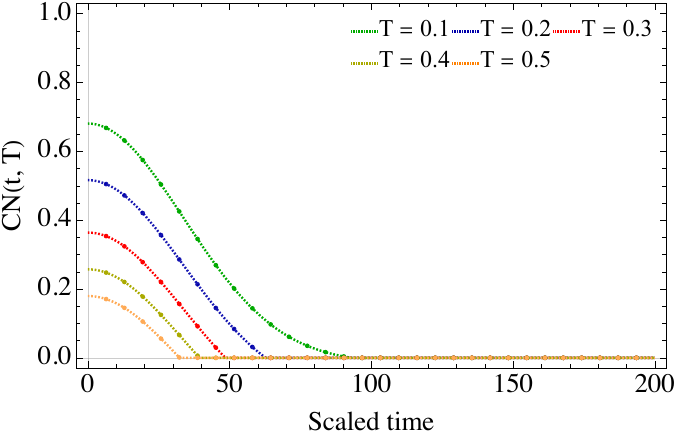}
		\put(-200,148){($ c $)}\quad
		\includegraphics[width=0.47\textwidth, height=145px]{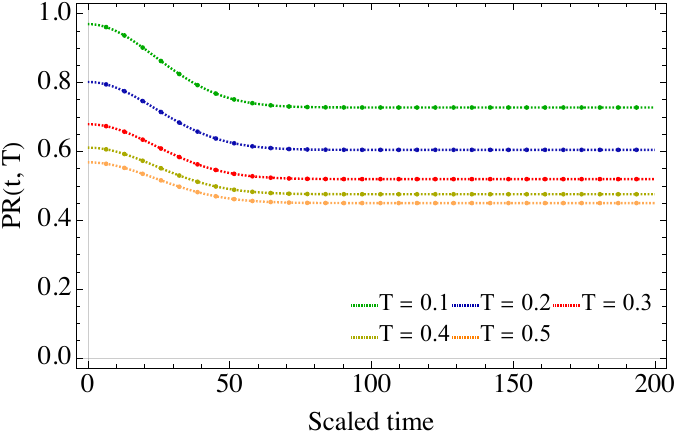}
		\put(-200,148){($ d $)}\quad
		\end{center}
\caption{Steerability (a), Bell non-locality (b), concurrence (c) and purity (d) are shown as in Fig. \ref{fig1}  but for fixed values of $T$ with $\omega =1$.} \label{fig4}
\end{figure*}

In Fig. \ref{fig4}, the impact of different fixed temperature values on the dynamics of gravcat steerability, non-locality, entanglement, and purity when jointly influenced by the PL noise is studied. The steerability and purity functions, ST and PR, seem well agree with each other and remain nearly maximal for the low temperature $T=0.1$. The non-locality and entanglement functions, BN and CN, remain non-maximal, even for the lower values of the temperature. Hence, the temperature is differently affecting the initial strength of two-qubit gravcat quantum correlations. Precisely, one can easily demonstrate that for the least temperature $T$ values, steerability and purity are maximal, however, for the increasing $T$ values, the initial robustness in the relative functions decreases. Most importantly, for the extreme temperatures ($T>0.2$), the BN completely vanishes, hence, remaining fragile compared to the steerability, entanglement, and purity in the gravcats. Furthermore, with time, the steerability always remains non-zero for various temperature values.
Note that the entanglement vanishes for all values of $T$ in long periods of time. Hence, there is no temperature gauging in the current configuration which could lead to the preservation of entanglement and specifically non-locality in the gravcat state. As a result, the phenomenon of complete freezing is observed for BN and CN but not for ST function.

\begin{figure*}[!t]
	\begin{center}
		\includegraphics[width=0.47\textwidth, height=145px]{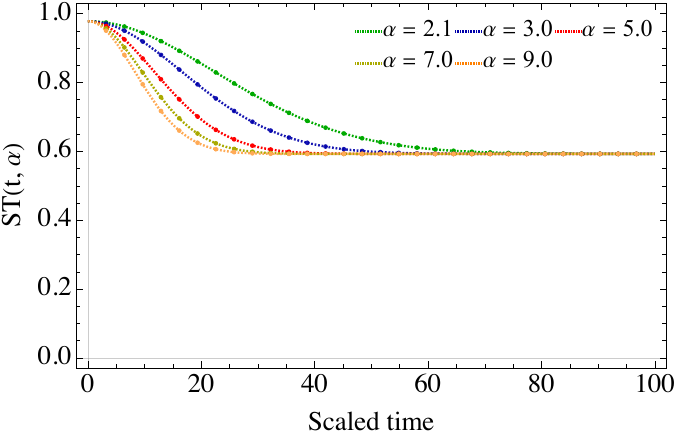}
		\put(-200,148){($ a $)}\quad
		\includegraphics[width=0.47\textwidth, height=145px]{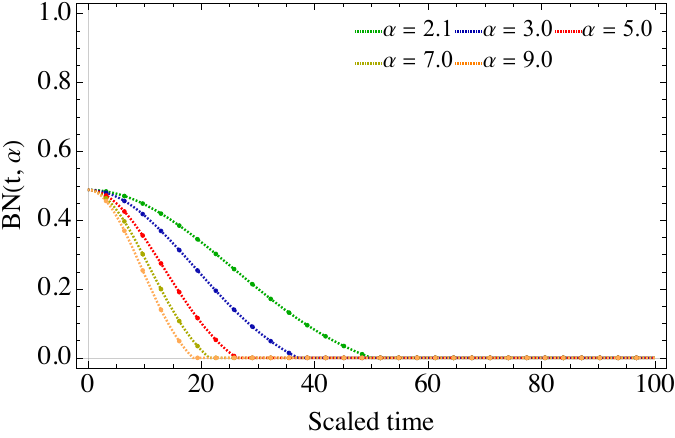}
		\put(-200,148){($ b $)}\\
		\includegraphics[width=0.47\textwidth, height=145px]{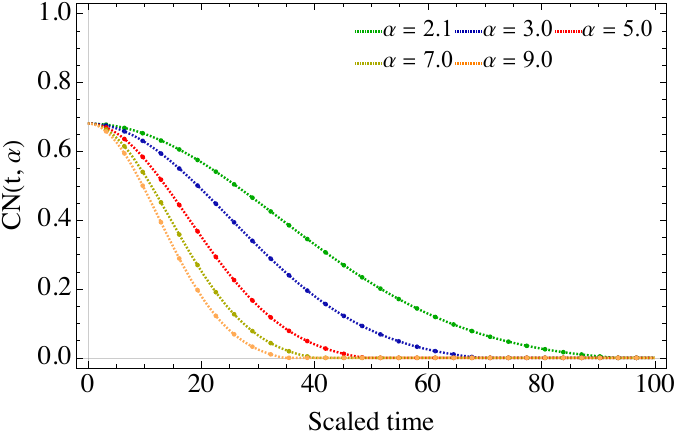}
		\put(-200,148){($ c $)}\quad
		\includegraphics[width=0.47\textwidth, height=145px]{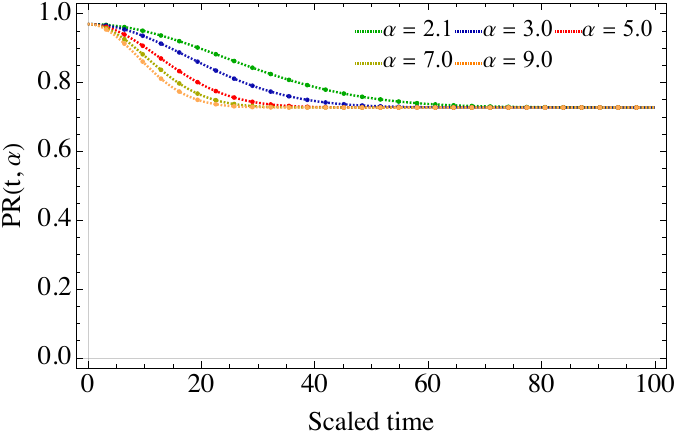}
		\put(-200,148){($ d $)}\quad
		\end{center}
\caption{Steerability (a), Bell non-locality (b), concurrence (c) and purity (d) are shown as in Fig. \ref{fig1}  but for fixed values of $\alpha$ with $\omega =1$.}		
\label{fig5}
\end{figure*}

Finally, the two-qubit steerability, non-locality, entanglement, and purity are depicted in Fig. \ref{fig5} versus scaled time and fixed values of noise parameter $\alpha$.
One can see that the two-qubit non-classical correlations decline with time for different values of $\alpha$ but at different rates and saturation levels. For example, the two-qubit steerability decreases at a faster rate for the higher $\alpha$ values and vice versa.
On the other hand,  BN and CN decline directly with the increasing $\alpha$ values and vanish completely at some specific interval of time, suggesting the less strengthened nature of the entanglement and non-locality in the gravcat state compared to the one-way steerability.  Moreover, the PR function provides a similar decay pattern and achieves a nearly maximal pureness limit after some duration for both higher and lower values of  $\alpha$.
Note that the dynamical map of the gravcat quantum correlations completely remains monotonic and no revivals are observed.
This exponential behavioral dynamics of the open gravcat system shows that it does not exchange information and loses the information permanently.

\subsubsection{Weak measurement protocol}
A quantum system is intrinsically open to its surroundings, making it susceptible to environmental noises. This will cause some dissipation and quantum decoherence in a quantum system. To reduce the attenuation of an idealized quantum system, which is recognized as a crucial component in quantum precision measurement, Aharonov \cite{Aharonov1988} introduced a ground-breaking technique called quantum weak measurement (QWM). With this in mind, we would like to see how well a specific operation can improve steerability, non-locality, entanglement, and purity while increasing system robustness. Fortunately, our investigation has yielded evidence that the solution is correct. We will go into more detail about how this operation relates to our problem in the paragraphs that follow.

First, we write the associated non-unitary operator for the QWM as \cite{21,QWM1}
\begin{align}
\mathcal{Q}=\left[\begin{array}{cccc}
 1 & 0\\
0 & \sqrt{1-r_d}
\end{array}\right],\label{QWM}
\end{align}
with the measurement strength $0\leq r_d \leq 1$, representing the operator acting on the particle $d$ (with $d\in\{A,~B\}$).

Then, we consider that both $A$ and $B$ will have these treatments. Finally, the QWM can be done by using the equation
\begin{equation}
\rho(t, T)^Q=\frac{(\mathcal{Q}_A \otimes \mathcal{Q}_B) \rho(t,T)(\mathcal{Q}_A \otimes \mathcal{Q}_B)^{\dagger }}{P_s},
\end{equation}
where $\rho(t,T)$ is given in Eq. \eqref{final density matrix3}, and $P_s={\rm Tr}\left[(\mathcal{Q}_A \otimes \mathcal{Q}_B)\rho(t,T)(\mathcal{Q}_A \otimes \mathcal{Q}_B)^{\dagger }\right]$ is the probability of success of the measurement.

\begin{figure*}[!t]
	\begin{center}
		\includegraphics[width=0.45\textwidth]{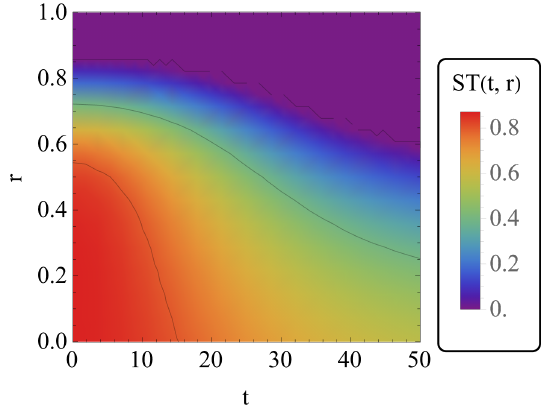}
		\put(-190,172){($ a $)}\quad
		\includegraphics[width=0.45\textwidth]{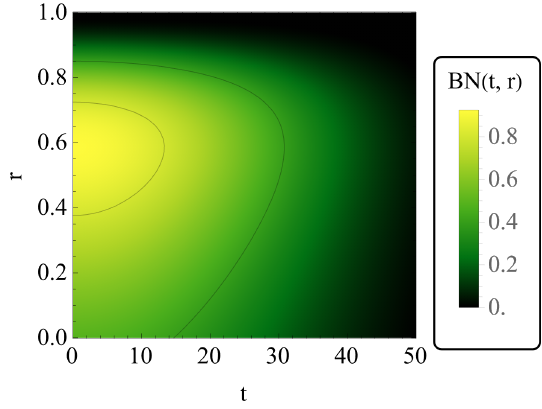}
		\put(-190,172){($ b $)}\\
		\includegraphics[width=0.45\textwidth]{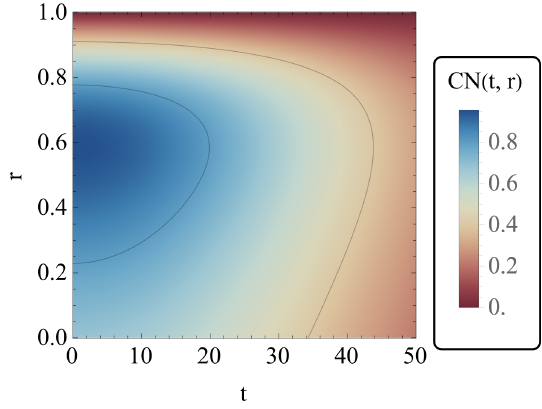}
		\put(-190,172){($ c $)}\quad
		\includegraphics[width=0.45\textwidth]{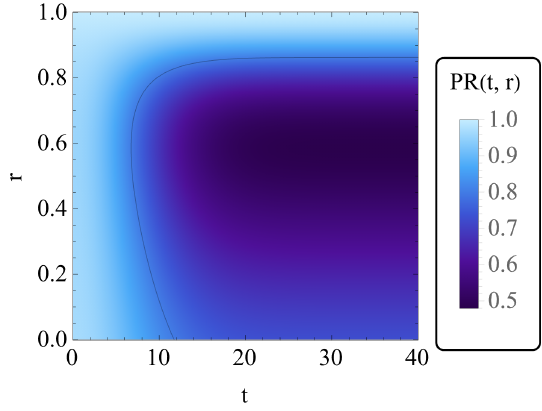}
		\put(-190,172){($ d $)}\
		\end{center}
\caption{Steerability (a), Bell non-locality (b), concurrence (c) and purity (d) as functions of $r$ and $t$ for two gravcats under weak measurement protocol.  For all plots  $g=10^{-4},~\gamma =1.5,~\omega=2,~\alpha=2.1,$ and $T= 0.1$.}\label{wmr}
\end{figure*}

For simplicity, we assume $r_d=r$, so the non-zero entries of the above matrix take the form
\begin{align*}
\rho_{11}^Q=&\frac{\rho_{11}}{P_s},\\
\rho_{14}^Q=&\rho_{41}^Q=\frac{e^{-8 \beta_{PL}(\tau)}\rho_{14}(1-r)}{P_s},\\
\rho_{22}^Q=&\rho_{33}^Q=\frac{\rho_{22}\sqrt{(1-r)^2} }{P_s},\\
\rho_{23}^Q=&\rho_{32}^Q=\frac{\rho_{23}\sqrt{(1-r)^2} }{P_s},\\
\rho_{44}^Q=&\frac{\rho_{44}(1-r)^2 }{P_s},
\end{align*}
where
$$P_s=2\rho_{22}\sqrt{(1-r)^2} +\rho_{44} (1-r)^2 +\rho_{11}.$$

In Fig. \ref{wmr}, we explore the impact of QWM on the time evolution of steerability, non-locality, entanglement and purity of the two-qubit state under PL noise. From the results, it is clearly predictable that the gravcat state can be successfully driven by QWM and certain values of $r$ are beneficial for the preservation of two-qubit correlations. However, it is noticeable that steerability, non-locality, entanglement and purity are differently affected by the QWM. For example, the steerability of the system remains robust for $0<r\lesssim0.8$, however, only for the initial time interval. Similarly, non-locality has also been demonstrated with the same behavior but it shows a clear maximum around $r\approx0.6$ initially. Entanglement has been found robust around  $r\approx0.6$ and reduced for all other values. The purity of the state seems maximum for all $r$ values. With time, the current two-qubit measures show a decline but with different rates. The steerability completely vanishes for $r>0.8$. The same is the case with non-locality, however, at the final notes of the time, even for all $r$ values, non-locality seems to dissipate completely. Likewise, it can be explored for entanglement and therefore shows strict agreement with non-locality of the system. Moreover, the purity decreases only in the range $0.2<r<0.8$.

\subsection{Extension of the model to electrostatic notion}
Interestingly, the structure of $\gamma$ given in Eq. \eqref{hamiltonian} is analogous to the electrostatic interaction coupling strength between two point charges $q$ given by the formula $k_\textmd{e}\frac{q^{2}}{2}\left(\frac{1}{\rm d}- \frac{1 }{{\rm d}'}\right)$ where $k_\textmd{e}=1/4\pi \varepsilon_0$ is the Coulomb constant with $\varepsilon_0$ denoting the vacuum permittivity. It is important to mention that, despite this similarity, there are also crucial differences between gravitational and electrostatic interactions. For instance, gravitational forces are always attractive and act between all masses, while electrostatic forces can be attractive or repulsive and act only between charged particles. Furthermore, the strength of the electrostatic force is much greater than that of the gravitational force, making gravity negligible on small scales compared to electrostatic forces.

Let us show a brief application of this notion for the Hamiltonian given in Eq. \eqref{hamiltonian}. In this regard, we assume a dipole moment vector with a magnitude of $|\mu|$. Here, the electrostatic interaction including the dipole moment is given by $\Gamma = k_\textmd{e} \frac{|\mu|^2}{d^3}$. Inserting this notion into the Hamiltonian  \eqref{hamiltonian}, we get

\begin{equation}\label{moment-hamiltonian}
\mathcal{H}_\text{dipole} = \frac{\omega}{2}\left(\sigma_{z}\otimes \mathbb{I}+\mathbb{I}\otimes \sigma_{z}\right) - \Gamma (\sigma_{x}\otimes \sigma_{x}),
\end{equation}
where $d$ is the separation between the qubit charges.

Now, we are dealing with a dimensionless parameter case; therefore, we omit the Coulomb constant $k_\textmd{e}$ in the above expression. { Afterwards, we prepare the thermal state of the above Hamiltonian using the Eq. \eqref{thermal density matrix1}. This thermal state is then exposed to the classical field influenced by PL noise using the methodology given in subsections \ref{classical} and \ref{power-law}. This would help us to demonstrate the basic differences between the gravitational and electrostatic state dynamics influenced by PL noise. The explicit final density matrix of the system has the form
\begin{align}
\rho^{ES}(t, T)=\left[
\begin{array}{cccc}
 \rho^{ES}_{11} & 0 & 0 &  e^{-8 \beta_{PL}(\tau)} \rho_{14}^{ES}\\[0.2cm]
 0 & \rho^{ES}_{22} & \rho^{ES}_{23} & 0 \\[0.2cm]
 0 & \rho_{23}^{ES} & \rho^{ES}_{22} & 0 \\[0.2cm]
 e^{-8 \beta_{PL}(\tau)} \rho_{14}^{ES}  & 0 & 0 & \rho^{ES}_{44} \\
\end{array}
\right].\label{rho-t2}
\end{align}
The entries of the above density matrix are given in Appendix \ref{appendixA}.

 Finally, by evaluating the final density matrix of the system $\rho^{ES}(t, T)$ with the electrostatic notion in terms of ST, BN, CN, and PR, we provide the detailed impact of the mentioned parameters on the preservation of quantum correlations.}

\begin{figure*}[!t]
	\begin{center}
	\includegraphics[width=0.35\textwidth]{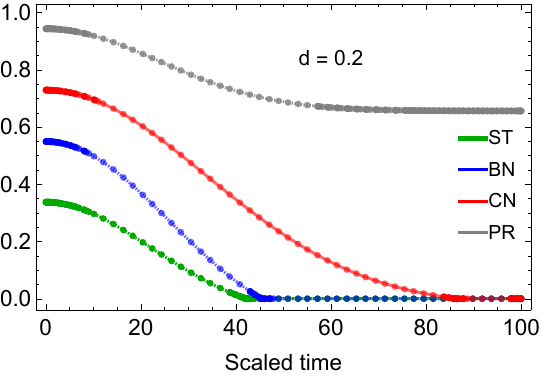}
		\put(-150,127){($ a $)}\quad
		\includegraphics[width=0.35\textwidth]{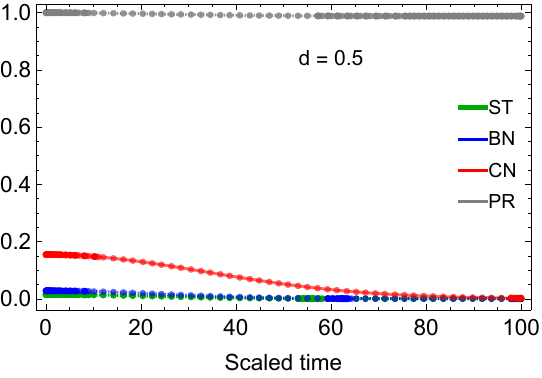}
		\put(-150,127){($ b $)}\\
				\includegraphics[width=0.35\textwidth]{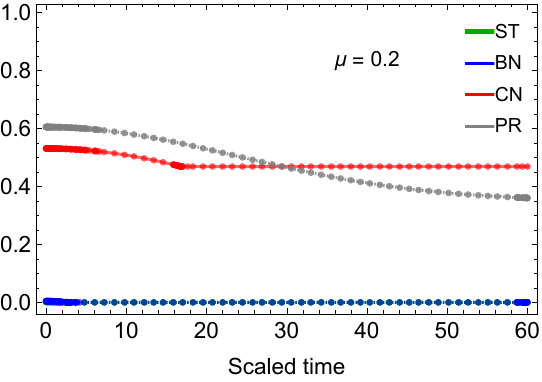}
		\put(-150,127){($ c $)}\quad
		\includegraphics[width=0.35\textwidth]{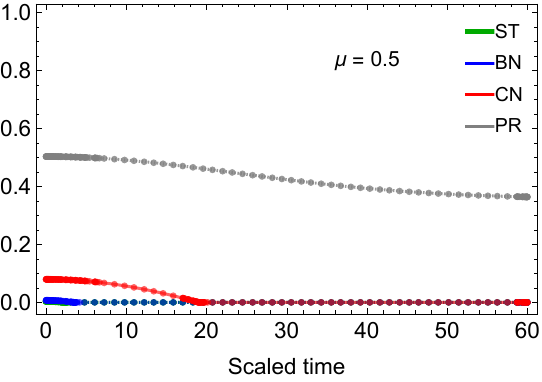}
		\put(-150,127){($ d $)}\\
		\includegraphics[width=0.35\textwidth]{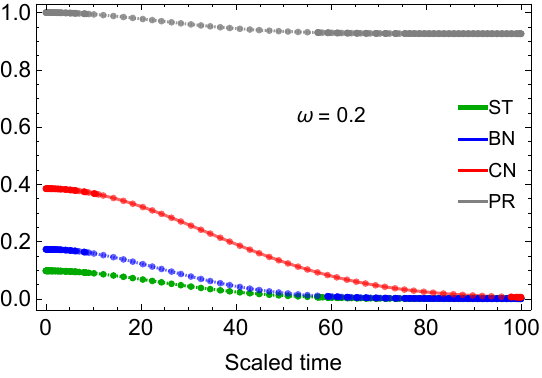}
		\put(-150,127){($ e $)}\quad
		\includegraphics[width=0.35\textwidth]{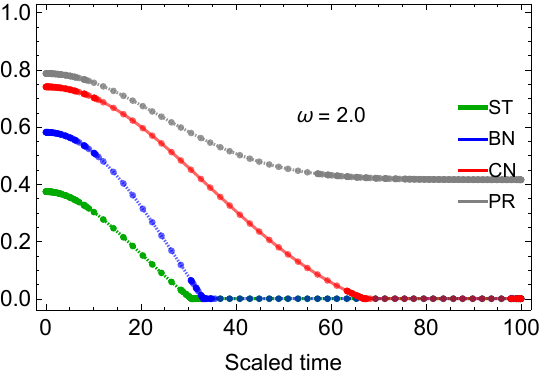}
		\put(-150,127){($ f $)}\\
		\includegraphics[width=0.35\textwidth]{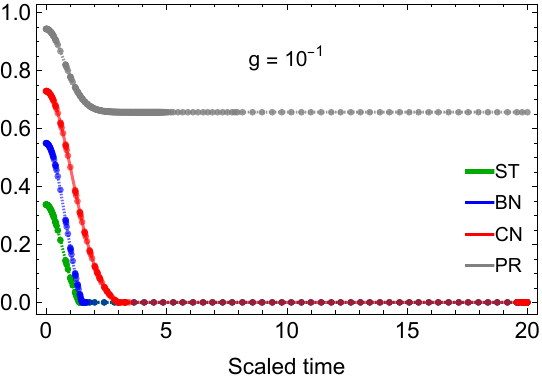}
		\put(-150,127){($ g $)}\quad
		\includegraphics[width=0.35\textwidth]{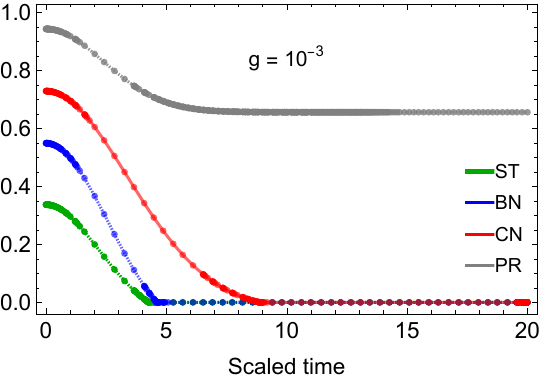}
		\put(-150,127){($ h $)}
		\end{center}
\caption{{  Steerability, Bell non-locality, concurrence, and purity:  (a, b)  as functions of scaled time and two fixed values of distance $d$ with $g=10^{-4},~\omega=1,~\alpha=2.1,$ and $T=\mu= 0.1$.  (c, d)  Same as above but for dipole moment $\mu$ with $g=10^{-4},~\omega=1,~\alpha=2.1,$ and $T=d= 0.1$. (e, f) Same as above but for the energy gap between ground and excited state $\omega$ with $g=10^{-4},~\alpha=2.1,$ and $T=d=\mu= 0.1$.  (g, h) Same as above but for noise field strength $g$ with $\omega=1,~\alpha=2.1,$ and $T=d=\mu= 0.1$.}}\label{ELS}
\end{figure*}
{  In Fig. \ref{ELS}, we particularly focus on replacing the gravitational notion with electrostatics in the Hamiltonian given in Eq. \eqref{hamiltonian}, which resulted in Eq. \eqref{moment-hamiltonian}. Here, we provide the analysis of the competency between the gravcat and electrostatic states to preserve quantum correlations. In Fig. \ref{ELS}(a, b), we characterize the quantum functions by $d$, and it is evident that the quantum functions remain non-maximal. Besides, the variation in $d$ seems not to help the quantum functions to remain preserved over time, therefore, complete destruction of quantum correlations occurs soon. It is crucial to note that under similar conditions, the gravcat state qubits were able to preserve ST the highest, while the electrostatic state seems to support purity, and entanglement the most, while non-locality and then ST seem enough sensitive. For the case of smaller, and larger $\mu$ against time in Fig. \ref{ELS}(c, d), ST and non-locality remained completely zero. However, purity and entanglement remain preserved for a while and then are lost completely.  When the quantum functions are characterized by $\omega$ in Fig. \ref{ELS}(e, f), we observe smaller initial values of quantum correlation functions, however, quantum correlations initially have high values for $\omega=2.0$ compared to the case of $\omega=0.2$. Therefore, they show an agreement with the results obtained for gravcat state when characterized by $\omega$ (Fig. \ref{fig1}). However, the net quantum correlation values and preservation time remain higher in the case of gravcat state compared to the electrostatic state. Finally, the impact of $g$ in Fig. \ref{ELS}(g, h) on the quantum correlation functions for the electrostatic state remains in agreement with that of the gravcat state (see Fig. \ref{fig2}). Overall, the current insights show that the gravcat state qubits seem more steerable as well as remain more preserved and suitable for quantum correlations preservation compared to the electrostatic state.}

{ 
\subsection{Impact of independent channels and non-uniform fields}
In the previous subsections, we assumed the exposure of the gravcat state to a single common channel where the field value does not vary. This assumption led to the off-diagonal element $\rho_{23}$ of the density matrix in Eqs. \eqref{rho-t} and \eqref{final density matrix3} not being affected by dephasing. Hence, the computed quantum features quantities show the preservation of quantum correlations even under the presence of dephasing. However, since the field is stochastic, and the two qubits are separated, therefore, it would be interesting to explore the dynamical behavior of the quantum features under independent channels. Let us assume two independent classical channels with damping, written as
\begin{align}
H(t)=&{H}_i (t)\otimes \mathbb{I}_j \otimes+\mathbb{I}_i \otimes{H}_j(t),
\end{align}
with
\begin{align*}
H_k (t)=E \mathbb{I}+\lambda \delta_k (t) \sigma_x=\left[
\begin{array}{cc}
 E  & \delta_k \lambda  \\
 \delta_k \lambda  & E  
\end{array}
\right],
\end{align*}
where $k \in \{i,j\}$. Here, we set $\delta_i \neq \delta_j$ in Eq. \eqref{ut} at the moment because we assume two independent fields with varying field strengths.  Next, using  \eqref{time evolved density matrix}, we obtain the time evolved state of the system as:
\begin{equation}
\rho(t, T)=U_i(t)U_j(t)\rho(0, T) U_i^{\dagger}(t)U_j^{\dagger}(t),\label{time evolved density matrix2}
\end{equation}
where $U_k(t)=\exp \left\{-i \int^{t}_{0} H_k (t^{\prime})dt^{\prime} \right\}$. 
Finally, we subject the case to the PL noise, using the dynamical expression in \eqref{final density matrix2}. However, as $\delta_i \neq \delta_j$, therefore, we set PL noise factors $\beta_i$ for qubit-$i$ and $\beta_j$ for qubit-$j$. Then, the mathematical form of the independent PL noise factor given in Eq. \eqref{beta function of PL} can be rewritten as:
\begin{equation}
\beta^{PL}_k(\tau)=\frac{g_k \tau(\alpha -2)-1+(1+g_k \tau)^{2-\alpha }}{(\alpha -2)g_k}.\label{beta function of PL2}
\end{equation}
Using now the PL noise factors, with the above-given condition and configuration, the final density matrix of the gravcat state influenced by two PL noise sources can be expressed as
\begin{equation}
\bar{\rho}(t, T):=\left\langle \left\langle U_i(t)U_j(t)\rho(0, T) U_i^{\dagger}(t)U_j^{\dagger}(t)\right\rangle_{{\eta_i(\tau)}} \right\rangle_{{\eta_j(\tau)}}.
\end{equation}

Afterwards, we present the results obtained, while we first compare the case where the impact of $g$ in Fig. \ref{fig2} for a common channel to the current configuration. Note that here two channels with independent noise parameters ($g_i$ and $g_k$) are involved compared to the previous case. 

\begin{figure}[!t]
	\begin{center}
	\includegraphics[width=0.42\textwidth]{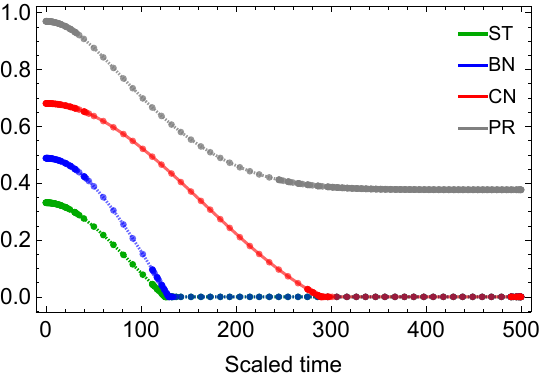}
		\end{center}
\caption{Steerability, Bell non-locality, concurrence, and purity against time for two-qubit gravcat state in independent fields influenced by PL noise with $g_i=g_j=10^{-5}$.  For all plots we fixed $\gamma=1,~\omega=1,~\alpha=2.1,$ and $T= 0.1$.}\label{fig12a}
\end{figure}

In Fig. \ref{fig12a}, we set $g_i=g_j=10^{-5}$ such that to find out how the gravcat state remained resourceful for quantum correlations compared to the case in Fig. \ref{fig2} with $g=10^{-5}$. We find the behavior of ST, BN, CN, and PR remains highly differing in the current case than observed in Fig. \ref{fig2}. For example, ST remained initially and at a later state preserved with higher values. In contrast, in the current case, ST remained sensitive initially and became lost easily with time. BN and CN show analogous behavior with almost similar preservation time approximately. The purity on the other hand remains a bit influenced negatively compared to the case in Fig. \ref{fig2}. Although, quantum correlations are lost after some time in the current case, however, it is still valuable to note that the preservation time obtained for the current quantum features in the state remained enough longer. This shows the resourcefulness of the gravcat state even when both the qubits are independently influenced by the external fields.

\begin{figure*}[!t]
	\begin{center}
	\includegraphics[width=0.445\textwidth]{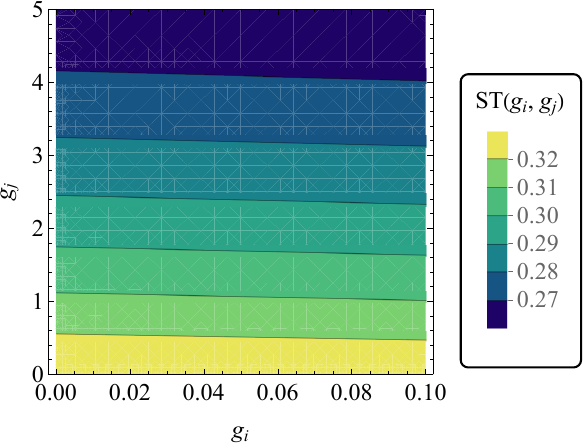}
		\put(-190,172){($ a $)}\quad
		\includegraphics[width=0.45\textwidth]{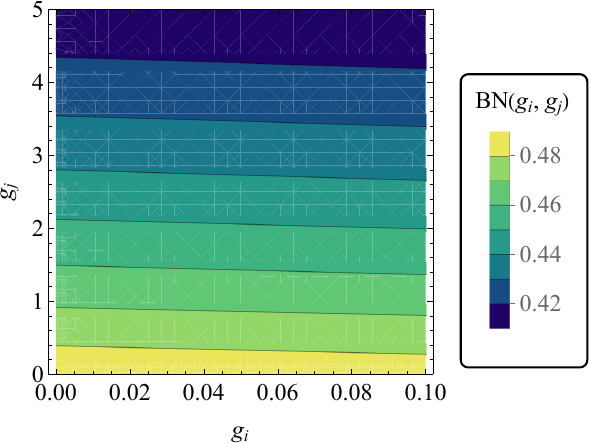}
		\put(-190,172){($ b $)}\\
			\includegraphics[width=0.445\textwidth]{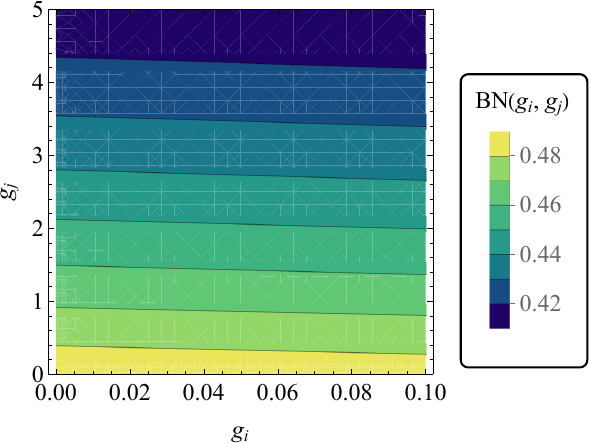}
		\put(-190,172){($ c $)}\quad
		\includegraphics[width=0.445\textwidth]{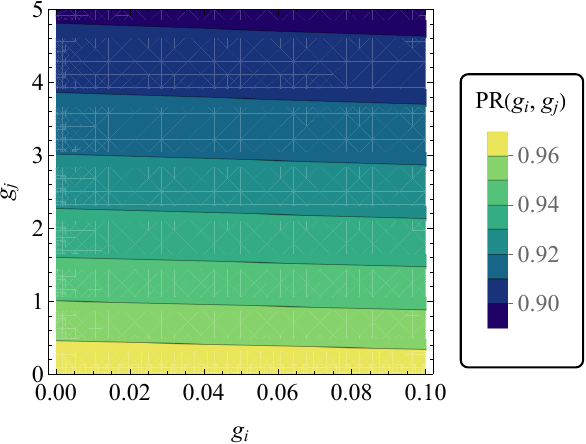}
		\put(-190,172){($ d $)}\\
		\end{center}
\caption{Steerability, Bell non-locality, concurrence, and purity as functions of parameter $g_i$ and $g_j$ for two-qubit gravcat state in independent fields influenced by PL noise.  For all plots we fixed $\gamma=\omega=t=1,~\alpha=2.1,$ and $T= 0.1$.}\label{fig12f}
\end{figure*}
In Fig. \ref{fig12f}, we investigate the impact of two-independent channels on the dynamical behaviors of quantum correlations in the gravcat state, specifically under the chosen ranges of $g_i$ and $g_j$. Compared to the previous case where the field parameter $\delta$ (given in Eq. \ref{inde ham}) was taken uniform, the steerability in the current case remains suppressed in both $g_i$ and $g_j$ ranges at the given time $t=1$. The BN, CN, and PR on the other hand remain less affected (compared to the case in Fig. \ref{fig2}). Therefore, the non-locality, entanglement, and pureness in the state remain strong even if the off-diagonal elements are influenced by the damping effects of the noise.  Hence, ST in comparison to the other existing quantum features turns out to be highly sensitive. However, this is interesting that ST can survive the best when a single common channel is introduced compared to two independent channels applied to the gravcat state. At the given time, the quantum correlations measurements show a minor decay without undergoing full destruction even if the noise field strengths are set to a stronger limit. Besides, it is crucial to note that if the independent qubit is exposed to a separate noise field and related strength, then measurements show higher destruction towards quantum correlations where the noise field strength is set to a stronger limit. For example, quantum correlations show higher decay against $g_j$ compared to $g_i$ because the variation range of $g_j$ is fifty times that of $g_i$.}

\subsection{Comparison of gravitational and electrostatic notions}
In this section, we intend to provide a comparative analysis of the gravcat and electric states under noisy conditions. Our motive is to differentiate clearly between the capacities of the two state to preserve quantum correlations. For this reason, let us do some normalization where we set $\tilde{\gamma} = \frac{\gamma}{\omega}$ for the gravcat state and the Hamiltonian in Eq. \eqref{hamiltonian} takes the following form after the normalization:
\begin{align}
  \frac{\mathcal{H}}{\omega} = &\frac{1}{2} \left(\sigma_{z} \otimes \mathbb{I} + \mathbb{I} \otimes \sigma_{z}\right) - \frac{\gamma}{\omega} \, (\sigma_{x} \otimes \sigma_{x}),\notag\\
  \tilde{\mathcal{H}} = &\frac{1}{2} \left(\sigma_{z} \otimes \mathbb{I} + \mathbb{I} \otimes \sigma_{z}\right) - \tilde{\gamma} \, (\sigma_{x} \otimes \sigma_{x}).
\end{align}
Here, the dimensionless coupling constant is defined as $\tilde{\gamma} = \frac{G m^2}{2 \omega} \left( \frac{1}{x} - \frac{1}{x'} \right)$ where \( G \) is the gravitational constant as before with \( G = 6.67 \times 10^{-11} \, \mathrm{Nm^2/kg^2} \). The inverse distance term is expressed as: $\frac{1}{r} = \frac{1}{x} - \frac{1}{x'}.$

Similarly, for the electrostatic model, we get
\begin{align}  \frac{\mathcal{H}_{\text{e}}}{\omega} = &\frac{1}{2} \left(\sigma_{z} \otimes \mathbb{I} + \mathbb{I} \otimes \sigma_{z}\right) - \frac{\Gamma}{\omega} \, (\sigma_{x} \otimes \sigma_{x}),\notag\\
  \tilde{\mathcal{H}}_{\text{e}} = &\frac{1}{2} \left(\sigma_{z} \otimes \mathbb{I} + \mathbb{I} \otimes \sigma_{z}\right) - \tilde{\Gamma} \, (\sigma_{x} \otimes \sigma_{x}).
\end{align}
Here, the dimensionless coupling constant is defined as $\tilde{\Gamma} = \frac{k q^2}{2 \omega} \left( \frac{1}{d} - \frac{1}{d'} \right)$ where \( k \) is the Coulomb constant with \( k = 8.99 \times 10^{9} \, \mathrm{Nm^2/C^2} \). The inverse distance term is expressed as: $\frac{1}{r} = \frac{1}{d} - \frac{1}{d'}.$ Note that the constants $G$ and $k$ are introduced to define the dimensionless coupling parameters $\tilde{\gamma}$  and $\tilde{\Gamma}$. While they carry physical units in their definitions, their role is to facilitate normalization, and the resulting Hamiltonians are entirely dimensionless, allowing for direct comparisons between the gravitational and electrostatic models.
{ Next, to make a realistic and experimental feasible comparison, we consider the two cases for the electrons trapped in the same double well potential, and placed at the same distance $r_e$.  Besides, we set the distance between the electrons fixed as $r=10^{-20}$ when treated as gravitational and electrostatic entities. Fixing this distance would provide an opportunity to observe the resourcefulness of the gravcat and electric state equivalently. Here, using the given fixed distance for the gravcat state, we get $\tilde{\gamma}=\gamma/\omega = G m_e^2/2r_e \omega = 6.09 \times 10^{-21}$ with $G=6.67\times 10^{-11}$ and $ m_e = 9.10 \times 10^{-31}$. For the electrostatic case, we get $\tilde{\Gamma}=\Gamma/\omega=k q_e^2/2 r_e \omega=1.44\times 10^{11}$ with $k=8.99 \times 10^9$ and $q_e=1.60218 \times 10^{-19}$}. Finally, we expose the above states to classical fields in Sec. \ref{classical}, and then apply PL noise given in Sec. \ref{power-law}.

\begin{figure}[!t]
	\begin{center}		\includegraphics[width=0.42\textwidth]{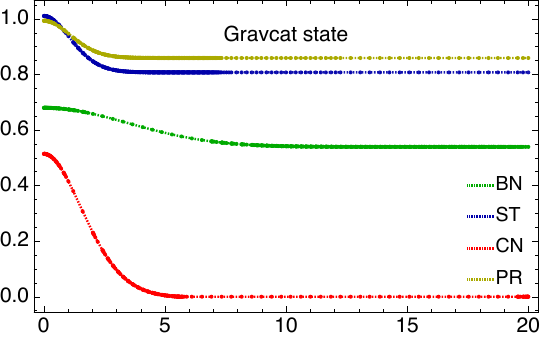}\put(-130,-5){ Scaled time}\\
\vspace{0.5cm}
\includegraphics[width=0.42\textwidth]{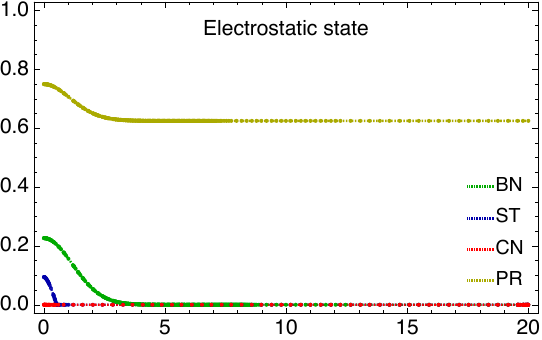}\put(-130,-5){ Scaled time}
	\end{center}
	\caption{{Upper panel: Steerability, Bell non-locality, concurrence, and purity against time for two-qubit gravcat state in independent fields influenced by PL noise with $g_{a,b} = 10^{-1}$ when $\alpha = 2.1 $, $T/\omega = 0.1$, and $\tilde{\gamma}=\gamma/\omega = 6.09 \times 10^{-21}$. Bottom panel: Same as above but for the electrostatic state with $\tilde{\Gamma}=\Gamma/\omega=1.44\times 10^{11}$.}}
	\label{figgt}
\end{figure}
{
In Fig. \ref{figgt}, the gravcat state (upper panel) and the electrostatic state (lower panel) are compared, and it is seen that they are both sensitive to noise, but to different degrees. In the gravcat state, ST and PR both have high initial values and do not degrade significantly with time, thus suggesting good preservation of steerable correlations and the purity of the state. BN has a lesser initial level in comparison and shows a slight drop before reaching a stable value. Note that both the initial and final values of ST and PR remain higher than those of BN. Entanglement remains the weakest initially and decreases gradually with time, finally vanishing completely.

On the other hand, the results obtained for the electrostatic state show that the state remains completely non-resourceful under the given conditions. For instance, except for purity, all the correlation measures remain suppressed or show non-significant initial values. The PR measure remains initially high and shows a smaller decline before reaching a stable value. BN, followed by the ST measure, remains highly suppressed initially and quickly vanishes. Most importantly, the state remains separable initially and at later intervals, with no signs of entanglement in the electrostatic state under the given conditions. Hence, the gravcat state shows a greater edge in resourcefulness under the given circumstances.}

{ \subsection{The implication of non-Markovian master equation}
In this section, we focus on imposing a non-Markovian noise on the gravcat system using a master equation introduced by Daffer {\it et al.} \cite{Daffer2014, Pinto2013}. With colored noise dephasing, the master equation of the system at time $t$ can be defined by
\begin{equation}
\varrho(t,T)=\kappa\mathbb{L}\rho (t,T),  \label{me}
\end{equation}%
where $\rho (t,T)$ is given in Eq. \eqref{final density matrix3} and $\mathbb{L}$ is the Lindblad super-operator describing the
dynamics of the system. The time-dependence is regulated by the integral part $\kappa$ as:
\begin{equation}
\kappa\Phi =\int_{0}^{t}J(t-t^{\prime })\Phi (t^{\prime })dt^{\prime },
\end{equation}%
where $J(t-t^{\prime })$ is the kernel function controlling the environment's memory. Next, to employ the master equation, let us consider the associated Hamiltonian, defined as \cite{Daffer2014,Pinto2013,Ali2014}
\begin{equation}
\mathbb{H}(t)=\omega _{i}(t)s_{i}^{z}+\omega _{j}(t)s_{j}^{z},  \label{Ht}
\end{equation}%
with $\omega _{m}$($m=\{i,j\}$) being the independent random variable
satisfying the statistic of a random telegraph (RT) noise signal: $\omega
_{m}(t)=a_{m}(-1)^{n_{m}(t)}$ in which $n_{m}(t)$ is a Poisson distribution
with a mean $\frac{t}{2\tau _{m}}$ and $a_{m}$ is a coin-flip parameter
having the values $\pm a_{m}$. We set $a_{i}=a_{j}=a$, and
$\tau _{i,j}=\tau $ for simplification. Using the stochastic Hamiltonian (\ref{Ht}), the correlation expression of the RT signal reads
\begin{equation}
\left\langle \omega _{m}(t)\omega _{m}(t^{\prime })\right\rangle =a^{2}\exp
(-\Gamma \left\vert t-t^{\prime }\right\vert ),
\end{equation}%
where $\left\langle \cdots \right\rangle $ denotes the ensemble
average while $\Gamma $ is the fluctuating rate of the RT signal.

In the case of the above master equation to be employed, various approaches are presented to examine the dynamical aspects of a quantum system \cite%
{Nielson2002,BenabdallahF2021} as well as the decomposition by Kraus operators \cite%
{Breuer2002,Ali2014}. Most specifically, it is more convenient to
utilize Kraus operators to investigate the dynamics of the gravcat state \cite{Breuer2002,Ali2014}. The Kraus operators associated with the above-described master equation with RT signals for the gravcat qubit-qubit system can be written as \cite{Daffer2014,Pinto2013}
\begin{equation}
\xi_{1}=\sqrt{\frac{1+\Lambda (\nu )}{2}}\mathbb{I}, \quad \xi_{2}=\sqrt{\frac{%
1-\Lambda (\nu )}{2}}s_{z},
\end{equation}%
while satisfying the normalization condition $\sum_{i}\xi_{i}^{\dagger }\xi_{i}=\mathbb{I}$, and $\mathbb{I}$ is the identity operator acting on the qubit space. $\Lambda (\nu )$ component of the above controls the harmonicity
of the oscillator resulting in both Markovian and non-Markovian noises, written as
\begin{equation}
\Lambda (\nu )=\frac{1}{\mu }\exp (-\nu )[\sin (\mu \nu )+\mu \cos (\mu \nu
)],
\end{equation}%
where $\mu =\sqrt{(4a\tau )^{2}-1}$ and $\nu =t/2\tau $. The final expression to superimpose the RT non-Markovian noise on the dynamical aspects of the gravcat system influenced by the thermal field driven by PL noise would be given by
\begin{equation}
\varrho (t,T)=\sum_{i,j=1}^{2}(\xi_{i}\otimes \xi_{j})\rho
(t,T)(\xi_{i}\otimes \xi_{j})^{\dagger }.  
\end{equation}%
The explicit form of the above density matrix has an X shape, namely
\begin{align}\label{master}
\varrho(t, T)=\left[
\begin{array}{cccc}
 \varrho_{11} & 0 & 0 &  \varrho_{14}\\[0.2cm]
 0 & \varrho_{22} & \varrho_{23} & 0 \\[0.2cm]
 0 & \varrho_{23}^* & \varrho_{22} & 0 \\[0.2cm]
 \varrho_{14}^* & 0 & 0 & \varrho_{44} \\
\end{array}
\right].
\end{align}
The density matrix elements are given in Appendix \ref{appendixB}.

\begin{figure*}[!t]
	\begin{center}
	\includegraphics[width=0.35\textwidth]{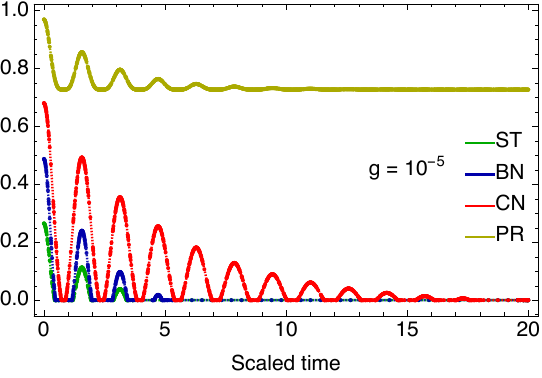}
		\put(-150,127){($ a $)}\quad
		\includegraphics[width=0.35\textwidth]{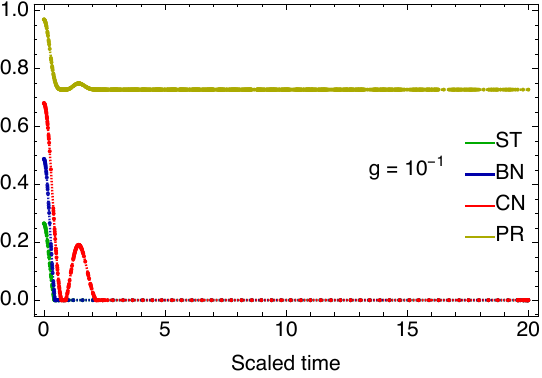}
		\put(-150,127){($ b $)}\\
				\includegraphics[width=0.35\textwidth]{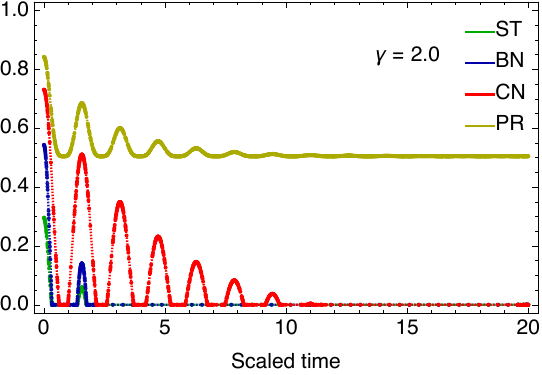}
		\put(-150,127){($ c $)}\quad
		\includegraphics[width=0.35\textwidth]{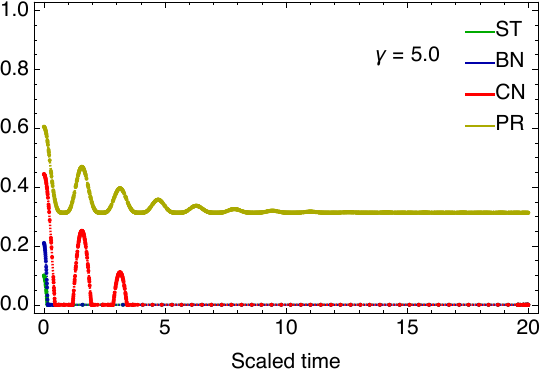}
		\put(-150,127){($ d $)}\\
		\includegraphics[width=0.35\textwidth]{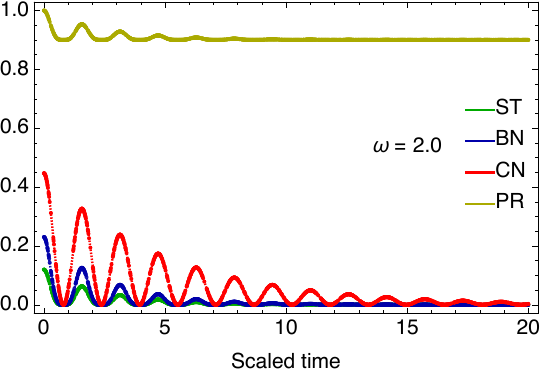}
		\put(-150,127){($ e $)}\quad
		\includegraphics[width=0.35\textwidth]{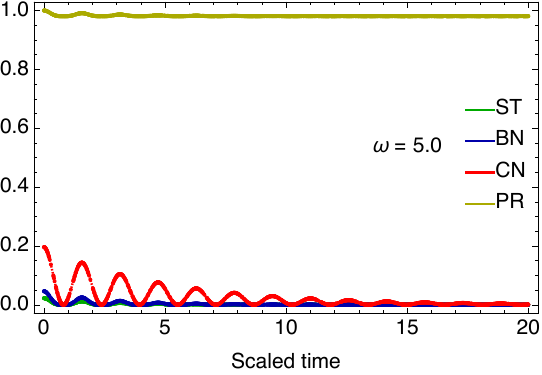}
		\put(-150,127){($ f $)}\\
		\includegraphics[width=0.35\textwidth]{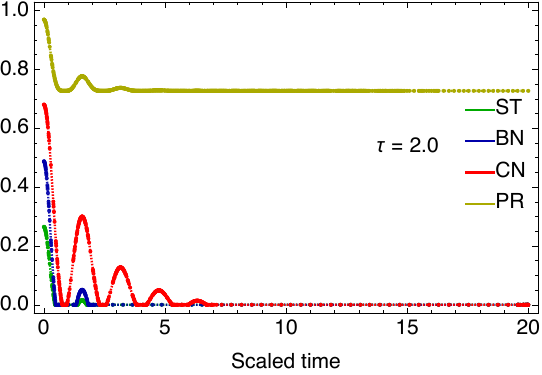}
		\put(-150,127){($ g $)}\quad
		\includegraphics[width=0.35\textwidth]{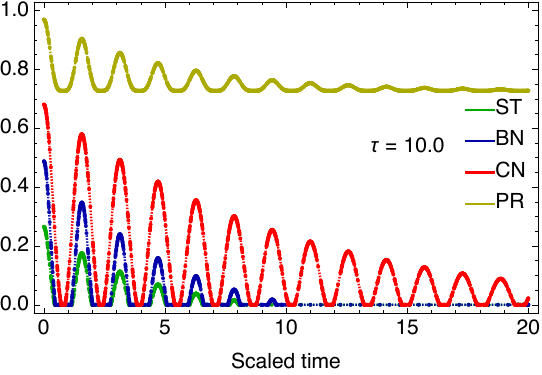}
		\put(-150,127){($ h $)}
		\end{center}
\caption{Steerability, Bell non-locality, concurrence, and purity: (a, b) as functions of scaled time and two fixed values of PL noise parameter $g$  with $\gamma=\omega=1,~\alpha=2.1,$ $\tau=5$.  (c, d) Same as above but for fixed values of $\gamma$ with $g=10^{-4},~\omega=1,~\alpha=2.1,~\tau=5$. (e, f)  Same as above but for fixed values of $\omega$ with $g=10^{-4},~\gamma=1~\alpha=2.1,$ $\tau=5$. (g, h) Same as above but for fixed values of non-Markovianity factor $\tau$ with $g=10^{-4},~\gamma=\omega=1,~\alpha=2.1$. Two fixed values are: $a=1$ and $T=0.1$.}\label{fig15}
\end{figure*}
Figure \ref{fig15} illustrates the dynamics of ST, BN, CN, and PR functions for the gravcat system when influenced by PL and RT noises. In the first case, we characterize the quantum functions by PL noise field strength $g$. The motive is to illustrate the competency between the Markov power of PL and the non-Markov power of the RT noise. 

In Fig. \ref{fig15}(a, b), at the onset, the quantum correlations functions seem non-maximal except for the purity. For the weaker coupling strengths of PL noise parameter $g=10^{-5}$, the RT noise traits become evident, causing sudden death and births of quantum correlations. However, for the increased $g$ values, i.e. $g=10^{-1}$, the non-Markovian trait only exists for the CN function for a shorter time while ST, and BN function undergo complete destruction. 

Figure \ref{fig15}(c, d) shows the impact of gravitational field strength $\gamma$ in the weaker and stronger coupling limits. For the smaller $\gamma$ value ($\gamma=2$), PR followed by CN measures remain preserved for a longer time. The CN function shows sudden death and births with revivals having higher amplitudes and then vanishes completely over time. the ST function remain suppressed the most without showing rebirth after the second loss. For the higher $\gamma$ strength ($\gamma=5$), the quantum correlations functions become highly vulnerable to the dephasing effects and becomes easily lost. 

When the functions are mainly defined by $\omega$ in Fig. \ref{fig15}(e, f), we observe smaller initial values of ST, BN, and CN functions when compared to the current cases of $g$ and $\gamma$. However, the named functions interestingly remain oscillating and preserved for longer duration in comparison. For the higher $\omega$ strengths, the ST, BN, and CN functions show smaller initial levels and the relevant revivals become suppressed with lesser amplitudes.
Surprisingly, purity seem highly enhanced and preserved for the gravcat system for the increasing $\omega$ strength. 

Finally, in Fig. \ref{fig15}(g, h), the case of non-Markovian factor $\tau$ is expressed. It is evident that the non-Markovian factor $\tau$ remains highly influential in the prolonged preservation of quantum correlations in the gravcat state. For example, the quantum correlations undergo fewer oscillations and disappear faster for $\tau=2$. However, as we increase $\tau$ to $10$, the oscillation in quantum functions gets enhance, and the preservation limit increases sufficiently. In comparison to the plots illustrated in Figs. (\ref{fig1})-(\ref{fig4}), we observe non-Markovian traits in comprehending the information feedback phenomenon (information exchange between the system and coupled environments). This is important for the gravcat state to reconvert from the classical regime to the quantum regime. One should note that the current configuration exhibits both the characteristics of Markov and non-Markov environments, and the prevalence of the traits displayed in the dynamics of the system depends upon the relative noise field strengths. 
Notably, we find the PL Markov noise [see Figs. (\ref{fig1})-(\ref{fig4})] more suitable for longer and higher preservation of quantum correlations than the RT noise dephasing. 

Overall, in the current case of RT noise, the gravcat state showed a valuable time interval to preserve quantum correlation, although smaller in comparison to PL noise. However, it should be noted that the gravcat state is subjected to the thermal field, PL noise, and finally RT noise. Yet, the gravcat state remained successfully preserved quantum correlations for a certain interval. Hence, the gravcat systems remain a reliable resource for quantum information processing compared to the simple and other types of open quantum systems, e.g., see Refs. \cite{rahmanadp, 36, EW, 35, 46, Ali2014} }

\section{Significance and experimental prospects of the study} \label{Significance}
The realization of open quantum system dynamics that exhibit optimal characteristics of longer information preservation is a hotly debated topic among quantum theorists and practitioners. In this regard, we presented a gravcat state coupled with a thermal-classical field. Previously, thermal and classical fields have been investigated extensively when considering their implementation in independent modes \cite{3, 34, 35, 36}. Both the classical and thermal fields have been found to completely decay the initially encoded quantum correlations in the system. As with quantum systems and channels, coupling is a sensitive process and might produce different results with slight changes.

Likewise, we observed that when jointly imposed on the gravcat state, the thermal-classical field remains more resourceful than that found in independent modes. However, the gravcats and thermal-classical field still exhibit parameter settings that should be completely avoided for the longer quantum correlations and information preservation. Besides, some studies have shown that simpler and thermal qubit-qubit states remain less resourceful than the currently employed gravcat state, for example, see Refs. \cite{34, 45, 46, 47}. Hence, this suggests that the gravcats are compelling candidates to be employed for practical quantum information processing.

Furthermore, the superposition of two mass particles under the influence of gravity has been proposed in Ref. \cite{48}. In this respect, they presented a phase evolution of two-micron size masses influenced by gravitational interaction using matter-wave interferometers, where they can be detected as entangled, even if they are detached at a large distance such that the Casimir-Polder forces are kept at the edge.  Marletto and Vedral proposed an experiment for witnessing the entanglement induced between two masses when impacted by the gravitational interaction \cite{49}. Despite these results, further rigorous investigations are needed to illustrate the influence of gravity and to employ it as a quantum resource.

Although the electrostatic and gravitational terms we considered have similarities in mathematical applications and some context, the impacts of both remain completely different. Like the gravitational system, the electrostatic system we considered can be proposed as a competitive candidate for the case of quantum gates and associated manipulation, as practically shown in Refs. \cite{gt1, gt2}. In a similar study, charge qubits are realized as coupled quantum dot chains, transforming into a register of transistor-like components, which may be an extension to aiding in the field of scalable quantum computers \cite{gt3}. Besides, the current-like electrostatic interaction remains involved in two-qubit transformations between electrons and adjacent dots while utilizing certain parameterizations \cite{gt4}.

In this study, we showed that non-classical correlations can be successfully modeled in the qubit-qubit gravcat state initially as well as for longer times. The one-way steering, BN, entanglement, and purity of the gravcat state all have been found to correlate with each other, showing prospects for experimentally detectable quantum correlations.

\section{Conclusion and outlook}\label{conclusion}
We presented the dynamics of two-qubit steerability, Bell non-locality, entanglement, and purity in a two-qubit gravcat state when coupled with the complex joint external environments consisting of thermal and classical fields. The initial level of quantum correlations in the gravcat state has been found dependent upon the state parameters. For example, when the gap between the ground state and excited state is increased, the degree of initially encoded quantum correlations becomes robust. In addition, for the strong gravitational interaction between the gravitating qubits, the associated quantum correlations were reduced initially as well as for the latter interval of time. The classical correlations in the gravcat state appeared for the increasing parameter values of the thermal-classical transmitting medium. 

The power-law (PL) noise parameters have been found to cause the decaying effects, however, without influence on the initial degree of quantum correlations in the state. In comparison, the gravcat state has been found more resourceful than the simpler two-qubit states, as the latter one exhibited complete decay in most of the external decoherence cases. Nonetheless, longer and larger preserved quantum correlations in the gravcat state have been detected using one-way steering, Bell non-locality, entanglement, and purity between them. 
Notably, the quantum correlations and purity are found to show similar qualitative dynamics mostly.  Moreover, overall behavioral dynamics remained monotonic and no revivals/fluctuations in any quantum criteria were recorded. 

Most importantly, the weak measurement protocol has been noticed to ensure the preservation of quantum correlations under certain conditions. However, know that the specific parameter related to this protocol needs to be set differently in the current model for the better preservation of different quantum correlation functions. Notice that the success probability of the measurement relies on the measurement strength, thus, quantum correlations are preserved at the expense of reducing the success probability of measurement. 

Besides this, we provided a brief overview of the extension of the gravitational model to an electrostatic model. In this case, we appraised the key difference between the relative parameters and their impact on the dynamics of quantum correlations. 

{Compared with similar field strengths, we found that the independent unequal noise strengths can cause higher destruction to quantum correlations. Moreover, superimposing non-Markovian factor over the Markovian PL noise acting on the gravcat state can bring relaxation to quantum correlations decay by causing the information exchange between the system and environment. It should be noted that we tested the gravcat state under thermal effect, PL Markovian noise, and a non-Markovian noise, yet the state remained successful in the preservation of quantum correlations, presenting itself as a resource.}

We demonstrated that gravcat states exhibit an exceptional degree of resilience against decoherence, rendering them a highly attractive contender for preserving quantum correlations across a broad spectrum of systems. The ramifications of this discovery are significant for the advancement of quantum technologies and our comprehension of the fundamental nature of quantum mechanics. Another area of research closely linked to gravcat states pertains to their potential in testing the underlying principles of quantum mechanics, particularly the concept of Bell non-locality. Given the entanglement of two massive objects separated by a macroscopic distance, gravcat states offer a unique channel for scrutinizing Bell non-locality. By measuring the correlations between these objects, we may potentially identify violations of local realism, thus providing further evidence for the peculiar and counterintuitive predictions of quantum mechanics. Furthermore, the exploration of gravcat states could divulge valuable insights into the relationship between quantum mechanics and gravity, which remains an intriguing and unresolved question in physics. By investigating the behavior of entangled massive objects in a gravitational field, we also may attain innovative perspectives on the nature of space-time and the fundamental forces governing our universe.

\appendix
\section{Elements of matrix $\rho^{ES}(t, T)$}\label{appendixA}
{
The entries of the density matrix \eqref{rho-t2} read
\begin{equation}
\rho_{11}^{ES}=\frac{\cosh \left(\frac{\sqrt{\mu ^4+d^6 \omega ^2}}{d^3 T}\right)-\frac{d^3 \omega  \sinh \left(\frac{\sqrt{\mu ^4+d^6 \omega ^2}}{d^3 T}\right)}{\sqrt{\mu ^4+d^6 \omega ^2}}}{Z^{\prime}},
\end{equation}
\begin{equation}
\rho_{14}^{ES}=\frac{ \mu ^2 \sinh \left(\frac{\sqrt{\mu ^4+d^6 \omega ^2}}{d^3 T}\right)}{Z^{\prime} \sqrt{\mu ^4+d^6 \omega ^2}},
\end{equation}
\begin{equation}
\rho_{22}^{ES}=\frac{\cosh \left(\frac{\mu ^2}{d^3 T}\right)}{Z^{\prime}},\label{electrostaic final}\\
\end{equation}
\begin{equation}
\rho_{23}^{ES}=\frac{\sinh \left(\frac{\mu ^2}{d^3 T}\right)}{Z^{\prime}},
\end{equation}
\begin{equation}
\rho_{44}^{ES}=\frac{\cosh \left(\frac{\sqrt{\mu ^4+d^6 \omega ^2}}{d^3 T}\right)+\frac{d^3 \omega  \sinh \left(\frac{\sqrt{\mu ^4+d^6 \omega ^2}}{d^3 T}\right)}{\sqrt{\mu ^4+d^6 \omega ^2}}}{Z^{\prime}}.
\end{equation}
where
$Z^{\prime}=2 \left[\cosh \left(\frac{\mu ^2}{d^3 T}\right)+\cosh \left(\frac{\sqrt{\mu ^4+d^6 \omega ^2}}{d^3 T}\right)\right]$.
}


\section{Elements of matrix $\varrho(t, T)$}\label{appendixB}
{
The gravcat system influenced by thermal field, and PL and RT noise configurations given in Eq. \eqref{master} has the following density matrix elements
\begin{equation}
\varrho_{11}=\frac{\left(\Lambda ^2+\sqrt{\left(\Lambda ^2-1\right)^2}+1\right) \left(\mathcal{K} \cosh \left(\frac{\mathcal{K}}{T}\right)-\omega  \sinh \left(\frac{\mathcal{K}}{T}\right)\right)}{2 Z \mathcal{K}},
\end{equation}
\begin{equation}
\varrho_{14}=\frac{e^{-8 \beta } \gamma  \left(\Lambda ^2-\sqrt{\left(\Lambda ^2-1\right)^2}+1\right) \sinh \left(\frac{\mathcal{K}}{T}\right)}{2 Z \mathcal{K}},
\end{equation}
\begin{equation}
\varrho_{22}=\frac{\left(\Lambda ^2+\sqrt{\left(\Lambda ^2-1\right)^2}+1\right) \cosh \left(\frac{\gamma }{T}\right)}{2 Z}, \label{non-markov append}
\end{equation}
\begin{equation}
\varrho_{23}=\frac{\left(\Lambda ^2-\sqrt{\left(\Lambda ^2-1\right)^2}+1\right) \sinh \left(\frac{\gamma }{T}\right)}{2 Z},
\end{equation}
\begin{equation}
\varrho_{44}=\frac{\left(\Lambda ^2+\sqrt{\left(\Lambda ^2-1\right)^2}+1\right) \left(\mathcal{K} \cosh \left(\frac{\mathcal{K}}{T}\right)+\omega  \sinh \left(\frac{\mathcal{K}}{T}\right)\right)}{2 Z \mathcal{K}},
\end{equation}
where $Z$ is defined in \eqref{thermal density matrix}.
}

\noindent \textbf{Acknowledgments:}
This work was supported in part by the National Natural Science Foundation of China (NSFC) under the Grants 11975236, 12235008 and by the University of Chinese Academy of Sciences. \\

\noindent \textbf{Data availability:} No datasets were generated or analyzed during the study.\\


\noindent \textbf{Competing interests:} The authors declare no competing interests.



\end{document}